\documentclass[preprint,showpacs,preprintnumbers,amsmath,amssymb,nofootinbib]{revtex4}
\usepackage{epsfig}
\usepackage{graphicx}
\usepackage{dcolumn}
\usepackage{bm}
\usepackage{threeparttable}
\usepackage{lipsum}  
\usepackage{slashed}
\usepackage[utf8]{inputenc}  
 \usepackage{float}

\def\beq{\begin{equation}}
\def\eeq{\end{equation}}
\def\eeqn{\end{equation}}
\newcommand\iden{\leavevmode\hbox{\small1\normalsize\kern-.33em1}}

\newcommand{\bea} {\begin{eqnarray}}
\newcommand{\eea} {\end{eqnarray}}

\newcommand{\nn}{\nonumber}

\newcommand{\nm}{\nonumber}

\newcommand{\vt}{\tilde v}
\newcommand{\yt}{\tilde y}
\newcommand{\qt}{\tilde q}
\newcommand{\Qt}{\widetilde Q}
\newcommand{\topt}{\tilde t}
\newcommand{\bt}{\tilde b}
\newcommand{\hc}{\mathrm{h.c.}}
\newcommand{\Ht}{\widetilde H}
\newcommand{\Lt}{\widetilde L}
\newcommand{\Z}{\widehat R}
\newcommand{\wt}{\tilde \omega}

\let\jnfont=\rm
\def\NPB#1,{{\jnfont Nucl.\ Phys.\ B }{\bf #1},}
\def\PLB#1,{{\jnfont Phys.\ Lett.\ B }{\bf #1},}
\def\EPJC#1,{{\jnfont Eur.\ Phys.\ Jour.\ C }{\bf #1},}
\def\PRD#1,{{\jnfont Phys.\ Rev.\ D }{\bf #1},}
\def\PRL#1,{{\jnfont Phys.\ Rev.\ Lett.\ }{\bf #1},}
\def\MPLA#1,{{\jnfont Mod.\ Phys.\ Lett.\ A }{\bf #1},}
\def\JPG#1,{{\jnfont J.\ Phys.\ G }{\bf #1},}
\def\CTP#1,{{\jnfont Commun.\ Theor.\ Phys.\ }{\bf #1},}
\def\JHEP#1,{{\jnfont JHEP \ }{\bf #1},}
\def\NPPS#1,{{\jnfont Nucl.\ Phys.\ Proc.\ Suppl.\ }{\bf #1},}
\def\CPC#1,{{\jnfont Comput.\ Phys.\ Commun.\ }{\bf #1},}
\def\CPL#1,{{\jnfont Chin.\ Phys.\ Lett. }{\bf #1},}
\def\APPB#1,{{\jnfont Acta\ Phys.\ Polon.\ B }{\bf #1},}
\def\PR#1,{{\jnfont Phys.\ Rept.\  }{\bf #1},}
\def\CHC#1,{{\jnfont Chin.\ Phys.\ C }{\bf #1},}
\def\RMP{\jnfont  Rev. Mod. Phys.}

\def\lsim{\raise0.3ex\hbox{$<$\kern-0.75em\raise-1.1ex\hbox{$\sim$}}}
\def\gsim{\raise0.3ex\hbox{$>$\kern-0.75em\raise-1.1ex\hbox{$\sim$}}}
\def\q_slash{\not{\hbox{\kern-2.1pt $q$}}}
\def\p_slash{\not{\hbox{\kern-2.1pt $p$}}}
\begin{document}

\title{\ \\[10mm] The Contribution of Charged Bosons with Right-Handed Neutrinos to the Muon g-2 Anomaly in the
Twin Higgs Models}

\author{Guo-Li Liu$^{1}$\footnote{Email address: guoliliu@zzu.edu.cn}, Ping Zhou$^{2,3,4}$\footnote{Email address: pzhou@nssc.ac.cn}  }
\affiliation{ $^{1}$School of Physics and Microelectronics, Zhengzhou University, Zhengzhou 450001, Henan, China   \\
$^{2}$ National Space Science Center, Chinese Academy of Sciences, Beijing 100190, China  \\
$^{3}$University of Chinese Academy of Sciences, Beijing 100140, China \\
$^{4}$Beijing Key Lab of Space Environment Exploration, Beijing 100140, China    }

\begin{abstract}
We examine the charged boson and right-handed neutrino contribution to the muon $g-2$ anomaly
in the twin Higgs models with the joint constraints of the Higgs global fit data,
the precision electroweak data, the leptonic flavor changing decay $\mu \to e\gamma $,
and the mass requirement of the heavy gauge bosons.
It comes out with the conclusion that some parameters
such as the coupling of charged Higgs to the lepton $ y_\mu$,
 the top Yukawa $y_t$, and
the heavy gauge boson coupling to the lepton $V_\mu$
are constrained roughly in the range of
$ 0.12\lesssim y_\mu\lesssim 0.4$, $0.4\lesssim y_t\lesssim 0.9$,
and $ 0.47\lesssim V_\mu\lesssim 1$, respectively.
\end{abstract}

\maketitle

\section{Introduction}

The long-standing puzzle, the muon anomalous magnetic moment $a_\mu \equiv (g-2)_\mu/2 $
was measured by the E989 muon $g-2$ anomaly experiment in Fermilab(FNAL)\cite{fnal-g-2}
and the Brookheaven National Laboratory(BNL)~\cite{BNL-bennett},
 \beq%
a^{\rm FNAL+BNL}_{\mu}=(11659206.1 \pm 4.1) \times 10^{-10}
\eeq%
which has a $4.2\sigma$ deviation from the prediction of the SM~\cite{fnal-g-2,g-2-SM} 
\beq
\Delta a_{\mu}^{\rm FNAL+BNL}=(25.1  \pm 5.9) \times 10^{-10}.
\label{au-exp-theo}
\eeq
It may hint the existing of the new
physics beyond the SM. The difference between the experimental data and the SM prediction
determines that there is room for new physics to live in.
Various new physics scenarios try to explain the muon
$g-2$ excess, for recent works, see e.g.
Refs. \cite{g-2-review1,g-2-review2,g-2-review4-1610.06587,g-2-review5,2002.12347,1811.04777,2106.04466}.

The so-called 
twin Higgs (TH) models can be realized by extending the SM with the discrete twin 
 symmetry\cite{litt-hier1} and are quite appealing.
Firstly, the discrete symmetry in them connects the SM fields with the extended ones.
Secondly, more importantly, the extended fields
 are uncharged under the SM gauge groups. That is to say, they will appear as singlets,
and "show" in upcoming experiments purely as missing energy, escaping from
the current constraints of the LHC to the new particles\cite{arXiv:hep-ph/0007265}.

In TH models, the extra charged gauge bosons and charged scalars\cite{1501.07890,1905.02203}
may have the leptonic and the quark flavor changing couplings,
which will contribute to muon anomalous magnetic moment.
Since there is not any signal of the twin top and twin gauge bosons in the experiments,
their masses may be very heavy, and the phenomenological signatures are suppressed.
So the main contribution to the muon anomaly, comes not from heavy twin top and twin gauge bosons, but
from the charged Higgs via its Yukawa couplings to SM quarks and leptons.

 However, TH models have encountered difficulties in cosmological considerations.
In the simplest realization of TH ~\cite{litt-hier1} models,
the twin particles will eventually transfer their entropies into those of SM photons and neutrinos,
which results in the deviation from the observations~\cite{neff_exp1,neff_exp2},
greatly increasing the value of $N_{eff}$, which is the effective number of (light) neutrino species\cite{1905.08798,1611.07975-neff_mth}.

Various modifications are proposed to reduce the $N_{eff}$ value in TH models. See e.g, Refs\cite{modif-mth1,modif-mth2,modif-mth3-1703.06884,modif-mth4,modif-mth5-1905.00861,1611.07975-neff_mth,modif-mth6,modif-mth7}.
%
 Type I seesaw~\cite{type I} is one of the mechanisms to provide tiny neutrino masses by involving the exchange of right-handed neutrinos,
and is also one to lower the effective degrees of freedom contributed by the twin sector when it is embedded into TH models~\cite{modif-mth3-1703.06884}.
Typical lepton-flavor-changing couplings will appear via this mechanism,
leading to interesting phenomenological consequences.

Flavor changing couplings can contribute to the muon anomalous magnetic moment
not only at the one-loop level, but also via the two-loop diagrams.
Since the quarks are much heavier than the leptons,
the phase space suppression may be surpassed by the mass enhancement,
so the two-loop Barr-Zee diagrams\cite{barr-zee} may contribute much more largely than that from the one-loop ones,
if only the Higgs bosons are not too light\cite{0909.5148,1912.10225}.
Thus, in the following, we will calculate the contributions from both the one-loop and two-loop levels.
We will examine the relevant parameter space of TH models by considering the
joint effects from the theory, the precision electroweak data,
 the 125 GeV Higgs signal data and the muon $g-2$ anomaly.


This paper is organized as follows.
In Sec. II we simply present TH models and the relevant couplings.
In Sec. III we discuss the rough constraints of the relevant parameters in  TH Models.
In Sec. IV, we calculate the muon anomalous magnetic moment $g-2$ at the one-loop level.
The analytic expressions for the two-loop muon $g-2$ anomaly and the Higgs global fit are given in Sec. V.
In Sec. VI we calculate the contributions of the Barr-zee Diagrams and the total constraints.
Finally, the conclusion is drawn in Sec. VII.

\section{The TH Models and the Relevant Couplings }

The hierarchy problem \cite{hier_prob1,hier_prob2,hier_prob3},
which is induced by the disparity between the electroweak scale and the
Planck scale, is one of the most outstanding problems in particle physics
(see, e.g., \cite{susy-pheno}).
Both supersymmetry \cite{susy1} and the compositeness
of the Higgs\cite{compo-scale}, even the strong breaking models\cite{tc-review},
 where the electroweak scale originates from a supersymmetry breaking scale or a
composite scale,
or even abandoning the idea of the Higgs as the origin of the electroweak symmetry breaking,
have attempted to address the issue.

However, in the new physics models mentioned above,
the solutions to the hierarchy problem
generically contain the top quark partners which have SM colored charge
with the mass at the electroweak scale.
Such particles have very rich phenomenological possibilities which are easily found at
the Large Hadron Collider (LHC). However, so far, there is not any signal of this kind of particles,
which constrains the their masses severely: typically larger than 1 TeV
\cite{parti-tev,1905.08798}.
To satisfy the bounds of this kind, the parameters in these theories
need to be fine-tuned to fix the electroweak scale.

One solution to the above problem is to assume that the Higgs mass is protected by a
$Z_2$ discrete symmetry that copies the SM particles,
so the new particles associated with this symmetry do not have the SM color charge,
which is called the twin Higgs(TH) models\cite{litt-hier1}.
This situation makes the new particle states much more difficult
to be produced and detected at the large hadron colliders.

In original TH framework, the SM particles and their 
copies are related by the discrete $Z_2$ twin symmetry.
To contain a residual custodial symmetry,
the global symmetry of the Higgs sector in the simplest realization
can be taken as $SO(8)$ or $SO(7)$ \cite{Barbieri:2015lqa,Batra:2008jy,1501.07890,1905.02203}.
The SM Higgs doublet is a part of the pseudo-Nambu-Goldstone bosons (pNGBs),
which arise from the spontaneously breaking of the global $SO(8)$($SO(7)$) symmetry into $SO(7)$($G_2$).
The neutral Higgs mass, under the joint action of the global symmetry and the discrete twin symmetry,
is protected from one loop quadratic divergence.

This mechanism can stabilize the Higgs mass up to the energy scale at the order of $5-10$ TeV,
and solve the so-called "little hierarchy" problem \cite{arXiv:hep-ph/0007265}.
A lot of models of this kind
\cite{litt-hier1,litt-hier2,litt-hier3,litt-hier4,litt-hier5,litt-hier6,litt-hier7,litt-hier8,litt-hier9,1711.05300},
have been proposed.
TH \cite{litt-hier1} realization is one of the best known examples.
After the twin Higgs obtains a vacuum expectation value $f$,
the SM Higgs has appeared as a pseudo-Nambu-Goldstone (pNG) boson,
preventing the Higgs mass from quantum
corrections up to the scale $\Lambda_{TH}\sim  4\pi f$
because of the twin 
symmetry between the top and the colorless top partner.

Ref.\cite{litt-hier1} gives the Higgs contributed by the new gauge bosons and extra fermions
and estimates the fine-tunings. For example, the contribution of the top partners
to the Higgs potential is given as
\beq
m^2_h =\frac{3}{8\pi^2} \frac{y^2 M^2}{M^2-y^2f^2}\left(
M^2ln\frac{m_{T_A}^2}{m_{T_B}^2} - y^2
f^2 ln\frac{m^2_{m_{T_A}^2}}{m_{t_B}^2}\right),
\label{mh2}
\eeq
where $y$ is Yukawa coupling, $yHQ_LT_R + h.c.$, and $m_{T_A}^2=M^2+y^2f^2$,
$m_{T_B}^2=M^2$, $m_{t_B}^2=y^2f^2$. The reference finally gives that,
for the limit $M \to \Lambda$, with $f = 800$ GeV, $\Lambda \sim 4\pi f \approx 10$ TeV, from Eq. \ref{mh2},
the Higgs mass is 166 GeV and the fine-tuning is $11\%$ (1 in 9), which is acceptable.
And the Higgs mass is 153 GeV and the fine-tuning is $31\%$ (1 in 3) for $f = 500$ GeV, $\Lambda  \approx 6$ TeV.
So for scale up to the order of $5-10$ TeV, the pseudo-Goldstone Higgs mass will be protected against radiative corrections,
and one can refer to the original paper \cite{litt-hier1} for details.

\subsection{The Charged Higgses and the Yukawa Couplings to the Third Generation in Twin Higgs Models  }
Since the minimal coset $U(4)/U(3)$ does not contain a residual custodial symmetry,
and in the non-linear case the twin mechanism is not realized in the gauge sector within this global group,
while $SO(8)/SO(7)$ prevents a large custodial breaking in composite models of the twin mechanism,
the global symmetry breaking pattern of the simplest original TH model can be $SO(8)/SO(7)$ or
$SO(7)/G_2$ case\cite{Barbieri:2015lqa,Batra:2008jy,1501.07890,1905.02203}.
Hence there are $7$ pNGBs after the breaking and $6$ of them are eaten by
the ordinary and twin gauge bosons,
 there will be only one neutral scalar left.

Besides the SM-like neutral Higgs, there would be charged Higgses.
Some TH models introduce extra scalars for different goals.
For example, to provide suitable neutrino masses
via couplings to the right-handed neutrinos,
a $SU(2)_L$ singlet charged scalar $S^+$ was introduced in Ref.\cite{1702.04399},
while Refs.\cite{modif-mth4, 
modif-mth5-1905.00861} add a new scalar $\phi$ to have
the similar effect in the couplings with the leptons.

Extra charged scalars may also appear in the particle list due to the enlarging breaking mode.
The aim of the economical breaking choices mentioned above is to keep the breaking smallest,
but it can be otherwise.
For example, it can be $SO(8)\to G_2$ or $SO(2N)\to SO(N)\times SO(N)$ \cite{2202.01228}
(the former $N$ is for SM sector and the latter, for twin sector).

In the followings, we assume the global breaking is  $SO(8)\to G_2$ as an example.
After the six gauge bosons obtain masses, the left eight PNGBs can write as $\Pi$ matrix
 \begin{equation}
\Pi= \sqrt{2} \pi^{\hat{a}} T^{\hat{a}}, \quad\quad\quad \hat{a}=1,\dots,7,
\end{equation}
where $T^{\hat{a}}$ are the broken generators, defined as
\begin{equation}
T^{\hat{a}}_{ij}= -\frac{i}{\sqrt{2}}t^{\hat{a}8}_{ij}, \quad\quad \hat{a}=1,\ldots,7.
\end{equation}
here $t_{ij}^{ab}= \delta^a_i\delta^b_j-\delta^a_j\delta^b_i$,
and $\pi^{\hat{a}}$ are the goldstone fields. 

Therefore, the $8$ scalar degrees of freedom may be packaged into an {\bf 8} of $SO(8)$
\begin{equation}\label{U}
\phi= \exp i \frac{\Pi}{f}.
\end{equation}
The first $4$ components would comprise the Higgs multiplet and the latter $N$ the Twin Higgs.
%
%
%
%
%
They can also be parameterized via the decomposition
$\mathbf{8} = (\mathbf{2},\mathbf{1},\mathbf{2}) + (\mathbf{1},\mathbf{2},\mathbf{2})$ under $SU(2)_L \times SU(2)_{\Lt} \times SU(2)_{\Z}$ as
\beq
\label{higgses}
(\mathbf{2},\mathbf{1},\mathbf{2}): \, H
= {f \over \sqrt{2}} \begin{pmatrix} \pi_2 + i \pi_1 \\ \pi_4 - i \pi_3 \end{pmatrix} \, , \quad
(\mathbf{1},\mathbf{2},\mathbf{2}): \, \Ht
= {f \over \sqrt{2}} \begin{pmatrix} \pi_6 + i \pi_5 \\ \sigma - i \pi_7 \end{pmatrix} \, ,
\eeq
where $\frac{f}{\sqrt{2}}(\pi_2 + i \pi_1)$ can identified as the charged scalar $H^\pm$.
In some situations, $\wt^\pm \equiv f (\pi_6 \pm i \pi_5)/\sqrt{2}$
and $\wt_0 \equiv f \pi_7$  can also be taken as the long-lived particles.

At low energies the representations of SO(8) of the third generation quarks $q_L$ and $\qt_L$ can be written as
\bea
Q_L \!\!\!&=&\!\!\! v_b b_L + v_t t_L = \frac{1}{\sqrt{2}}
\begin{pmatrix}
i b_L & b_L & i t_L & - t_L & 0 & 0 & 0 & 0
\end{pmatrix}^T \,, \nn \\
\Qt_L \!\!\!&=&\!\!\! \vt_b \bt_L + \vt_t \topt_L = \frac{1}{\sqrt{2}}
\begin{pmatrix}
 0 & 0 & 0 & 0 & i \bt_L & \bt_L & i \topt_L & - \topt_L
\end{pmatrix}^T \,,
\label{topembed}
\eea
while those of $t_R$ and $\topt_R$ are singlets.
Thus the top Yukawa couplings are written as
\beq
y_t f \bar t_R \Sigma^\dagger Q_L + \yt_t f \bar \topt_R \Sigma^\dagger \Qt_L + \hc = - y_t \bar q_L H t_R  - \yt_t \bar \topt_R \Ht \qt_L + \hc +...\, ,
\label{yuk}
\eeq
where $q_L=(b,t),~\tilde{q}_L =(\tilde{b},\tilde{t})$.

A $Z_2$ symmetry $q_L, t_R \leftrightarrow \qt_L, \topt_R$ leads to $y_t = \yt_t$.
The masses of the rest fermions (including twin fermions) are obtained in a similar manner.
Due to $\yt_\psi \ll y_t$ the associated contribution to the Higgs potential of the light fermions
will be negligible,
so it is not needed to enforce the approximate equality $y_\psi = \yt_\psi$.

\subsection{Flavor Changing Couplings of Leptons in TH Models  }
In the seesaw type-I model~\cite{type I}, to realize the seesaw mechanism,
right-handed neutrinos are introduced, and they are singlets under the SM gauge group $SU(3)_C \times SU(2)_L \times U(1)_Y$ \cite{1006.5534}.
We denote  $\nu_i$s as the ordinary tiny neutrinos, and $\nu_{iR}$s as the heavy right-handed neutrinos.
The masses of the neutrinos can be described by the following Lagrangian terms
\beq
{\cal L}\supset  - Y_j \bar\nu_R  H^+p_L\ell_j- y_j \bar\nu  H^+p_L\ell_j + {\rm h.c.} +...,
\label{lfv-2}
\eeq


In the mass eigenstate basis, the gauge interactions with $\nu_R$ are given by\cite{Emam:2007dy,1105.1047},
\bea \nm
{\cal L} &\supset &-\frac{g}{\sqrt{2}} \left( \bar l_{jL}\gamma^\mu V_{PMNS}\nu_{iL}W^-_\mu + \bar l_{jL}\gamma^\mu V_{l\nu} (\nu_{iR})_LN W^-_\mu + h.c. \right) \\
&\to& V_{j}\bar\nu_R  W_{\mu}^{+}\gamma^\mu P_L \ell_j +h.c. + ...
\label{lfv-1}
\eea


Note that in the second lines of equations (\ref{lfv-2}) and (\ref{lfv-1}) and following calculation,
we assume that the right-handed neutrinos are degenerate, i.e, $\nu_{1R}=\nu_{2R}=\nu_{3R}=\nu_R$,
which means that there is only one flavor of the heavy neutrino (the same case for the ordinary neutrinos),
so the couplings 
will be simply written as $V_j$, $Y_j$ and $y_j$, respectively,
where $j$ can be $e$, $\mu$, and $\tau$.

Note that the $3 \times 3$  Maki-Nakagawa-Sakata (MNS) matrix $U_{MNS}$ \cite{mns-maki-1962,0712.4019}
elements in above flavor changing couplings are also absorbed into the couplings
and their effects are actually neglected.

\section{The rough ranges of the Relevant Parameters in TH models}

In this paper, we take the light CP-even Higgs $h$ as the SM-like Higgs, $m_h= 125.5$ GeV.
The mass difference between the charged
Higgs boson and the neutral heavy Higgs boson should
be less than $300$ GeV, i.e., $|m_{H^\pm}-m_0| \leq 300$ GeV,
and in general, the mass of the neutral heavy Higgs boson is larger than that of the SM-like Higgs \cite{charged-neutral-mass}.
We here, however, just roughly scan over the charged Higgs mass $m_{H^\pm}$ in the following ranges:
\beq
100 ~{\rm GeV} <m_{H^\pm}< 1000~ {\rm GeV}.
\eeq

In the following, we will discuss simply the constraints on the parameters:
\begin{itemize}
\item[(1)]  The first constraint comes from the signal data of the 125 GeV Higgs, which is important, since the
couplings of the SM-like Higgs with the fermions and the bosons in TH models can deviate from
the SM largely and the production and decay modes of the SM-like Higgs may be modified severely.
In the paper, we will perform the calculation of $\chi^2_h$ for the signal strengths of the 125 GeV Higgs, which
will be shown in Sec. V. and  Sec. VI.
\item[(2)]
The constraints on parameter $f$ come from the joint effects of the $Z$-pole precision measurements, the low energy
neutral current process and the high energy precision measurements off the Z-pole indirectly,
and according to all these data, $f$ should be larger than 500-600 GeV \cite{0611015-su}.
On the other hand, to control in a mild fine tuning,
 $f$ should not be too large since the fine tuning is more severe for large $f$.
The constraints for $f$ can also come from the flavor changing decay $\mu\to e\gamma $: 
With the experimental constraints Ref.\cite{MEG-2013-2016}, ${\rm{BR}}(\mu \to e \gamma) < 4.2 \times 10^{-13}$,
the flavor changing decay $\mu\to e\gamma $ will give $f \sim[ 0.6 \ -\  2]$~TeV.

\hspace{0.2cm} So after we take the above constraints from the electroweak precision measurements and the LHC data into account,
we can assume that $ 500 \leq f \leq 2000$ GeV. 
  In our numerical evaluations, however, we have not taken $f$ as free parameter.
Instead, we assume the characteristic mass and coupling of the composite resonances
 is set by $m$ and $g$ respectively, which are related by the symmetry-breaking order parameter, $f$, as $m = g f$.
\item[(3)] In the nowadays experiments, $m_{W^\pm_H}$ has been constrained stringently \cite{gau-bo-mass-1,gau-bo-mass-2,gau-bo-mass-3}.
The ATLAS experiment has presented the first search for dilepton resonances
based on the full Run 2 data set \cite{gau-bo-mass-1,gau-bo-mass-3}
and set limits on the $W'$ production cross sections times branching fraction in the process
\beq
\sigma(pp \to W'X )\times BR(W'\to \nu\ell)
\eeq
 for $M_W'$ in the $0.15$ TeV $-7$ TeV range, correspondingly. Recently,
similar searches have also been presented by the CMS Collaboration using $140$ $fb^{-1}$ of data
recorded at $\sqrt{s} = 13$ TeV \cite{gau-bo-mass-2}. The most stringent limits on the mass of $W'$
boson to date come from the searches in the above process by the ATLAS and CMS collaborations
using data taken at $\sqrt{s} = 13$ TeV in Run 2 and set a $95$\% confidence level (CL) lower limit
on the $W'$ mass of 6.0 TeV \cite{gau-bo-mass-3}.

\hspace{0.2cm}This analysis, however, is based on the simplest
models\cite{pdg-2018} such as the sequential standard model proposed by Altarelli et al.,\cite{ssm-1989},
which is usually taken as a convenient benchmark in the experiments.
In the simplest models, the gauge particles are considered as the copies of the SM gauge bosons,
and their couplings to fermions are in the same mode as those of the SM gauge bosons,
but they miss trilinear couplings such as $W'W Z$ and $Z'WW$, etc.
So the situation that the sequential standard model\cite{ssm-1989} has acted as a reference for experimental extended gauge
boson searches may be changed and the results may be re-interpreted in the context of other new physics models~\cite{1912.02106}.
In the following computation, we will check the
sensitivity of the charged heavy gauge boson with the mass range $1 \leq m_{W^\pm_H}\leq 20$ TeV.

\item[(4)] About the top Yukawa $y_t$,
at the EW scale, it should be the same as that in the SM, but at the higher scale, the twos will be different.
Since in general,
we assume that the top quark is connected to electroweak symmetry breaking and
sensitive to the new physics models, we scan the top Yukawa $y_t$ from zero to $1.5$ times
of the SM top Yukawa $y_t^{SM}$.
The heavy gauge boson couplings to the lepton $V_\mu$ is
also from zero to $1.5$ times of the SM couplings $V_\mu^{SM}$.

\hspace{0.2cm}From the relationship of the masses of the ordinary neutrino and the right-handed neutrino, $m_\nu \sim \frac{Y_\mu^2 v^2}{m_{\nu_R}}$~\cite{1611.07975-neff_mth}, we can estimate the Yukawa coupling $Y_\mu$
is roughly $10^{-5}-10^{-3}$ when the right-handed neutrino masses are taken in the order of TeV
and the ordinary neutrino masses are assumed as $10^{-3}-10^{-1}$ eV.
Or, if the right-handed mass $m_{\nu_R}$ are free, this relation may serve as a constraint on $Y_\mu$.
\end{itemize}

\section{ The one-loop Muon Anomalous Magnetic Moment $g-2$ }

\begin{figure}[H]
\centering
\includegraphics[width=10cm]{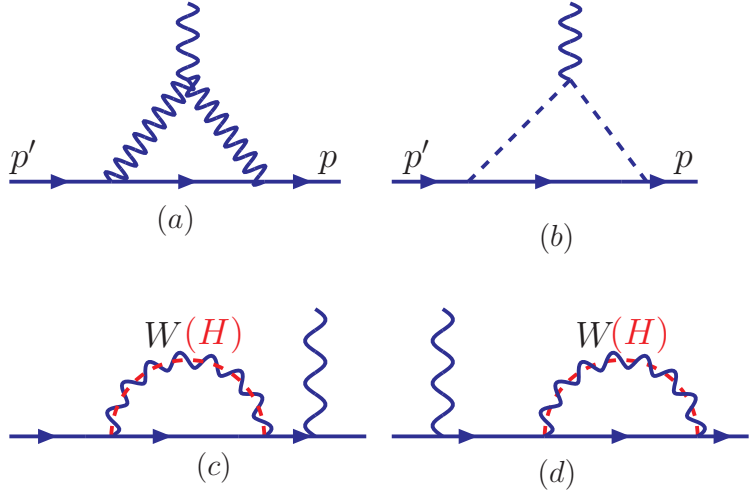} \vspace{-1.5cm}
\caption{The triangle and the penguin type diagrams for
the muon anomalous magnetic moment at the one-loop level.
The solid lines, wavy lines and dash lines denote the fermions, the gauge bosons and the charged Higgs, respectively,
which are the same as those in Fig.\ref{fig-2-loop}.}
\label{fig1}
\end{figure}

From the couplings of  Eq.(\ref{lfv-2}) and Eq.(\ref{lfv-1}), one can easily calculate
the new contributions to the muon anomalous magnetic moment which
are generated by one loop diagrams involving the exchange of
$W_{\mu}$ gauge boson and charged scalar $H^\pm$ with heavy neutrino,  as shown in Figure\ref{fig1}.

Note that the contributions from Fig.\ref{fig1}(c)(d)
contain self energy of the fermions, which is proportional to $\gamma^\lambda$,
and they do not contribute to the muon anomalous moment.
That is because,
when the Lorentz structure is written as the matrix element of the electromagnetic current
between incoming and outgoing fermion states of momentum and spin $\{p,s\}$ and $\{p',s'\}$, respectively,
\beq
<p,s|J_\lambda^{em}|p',s'>= \bar \mu_e(p,s)\{F_1(Q^2)\gamma_\lambda +\frac{F_2(Q^2)}{2m}\cdot\frac{i}{2}[\gamma_\lambda ,\gamma_\nu] \cdot q^\nu \} \mu_p(p',s'),
\label{lor_llr}
\eeq
the second term, $F_2(0)=\frac{g-2}{2}\equiv a_\mu$ ($Q=p'-p$ ),
while the first term, i.e, Fig.\ref{fig1}(c)(d) does not contribute the muon anomalous magnetic moment.
%

The one-loop contribution to the muon anomalous magnetic moment can be written as\cite{1loop-deltaamu}: 
\beq %
\Delta a_{\mu}^{TH}=\Delta a^{\nu_R W}_{\mu}+\Delta a^{\nu_R H}_{\mu}+\Delta a^{\nu H}_{\mu}+\Delta a^{\nu W_H}_{\mu}.
\eeq
 \beq
    \Delta a_\mu^{\rm \nu_R W}({\rm 1-loop}) =
    V_\mu^2 \frac{ m_{\mu}^2}{8 \pi^2 } \int_0^1 dx \frac{x^2(1+x)-x(1-x)\frac{m_{\nu_R}^2}{m_W^2}}{m_{W}^2 x+ m_{\nu_R}^2(1-x)},
\label{anu-w1}
\eeq
\beq
    \Delta a_\mu^{\rm \nu W_H}({\rm 1-loop}) =
    \left(\frac{g}{\sqrt 2}\right)^2 \frac{ m_{\mu}^2}{8 \pi^2 } \int_0^1 dx \frac{x^2(1+x)}{m_{W_H}^2 x+ m_{\nu}^2(1-x)},
\label{anu-w2}
\eeq
\beq
    \Delta a_\mu^{\rm \nu_R H}({\rm 1-loop}) =
    Y_\mu^2 \frac{ m_{\mu}^2}{16 \pi^2 } \int_0^1 dx \frac{x^3-x^2}{m_H^2 x+ m_{\nu_R}^2(1-x)},
\label{anu-h1}
\eeq
 \beq
    \Delta a_\mu^{\rm \nu H}({\rm 1-loop}) =
    y_\mu^2 \frac{ m_{\mu}^2}{16 \pi^2 } \int_0^1 dx \frac{x^3-x^2}{m_H^2 x+ m_{\nu}^2(1-x)},
\label{anu-h2}
\eeq
where $\frac{g}{\sqrt{2}}$ is the $W(W_H)$ boson couplings to the leptons, same as that in the SM.

\begin{figure}[H]%
 \begin{center}      
  \parbox{10cm}{\epsfig{figure=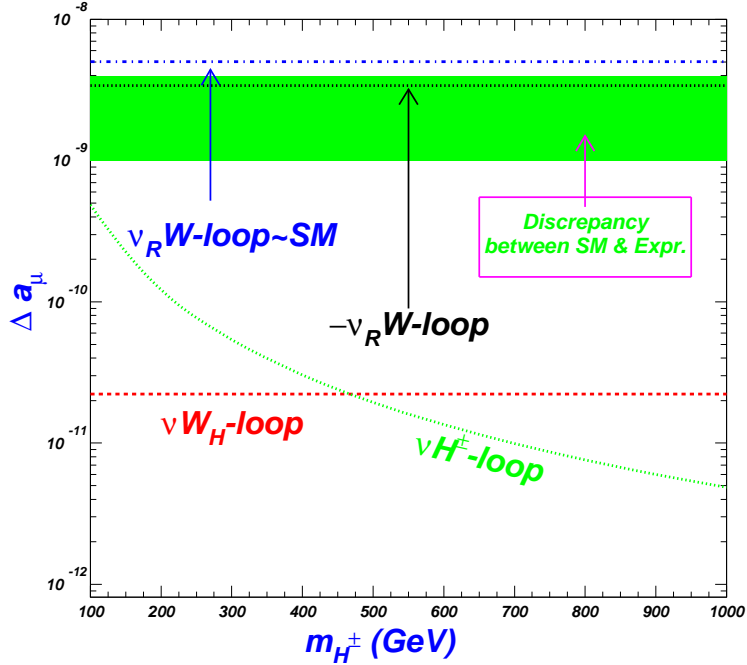,width=10cm}  }
 \caption{ The one-loop muon anomalous magnetic moment for
$V_\mu =0.6$, $Y_\mu=0.4$, $y_\mu=0.4$, $y_t=0.5$.
The green shadow area is the discrepancy between the SM and the measurement for the anomalous magnetic moment $\Delta a_\mu$.
 \label{fig2} }
   \end{center}
 \end{figure}

Fig.\ref{fig2} shows that the contributions of the charged Higgs, the heavy charged bosons and the heavy neutrino
at one-loop level, and we find that the contributions from $\nu W_H,~\nu H^\pm$ loop are quite small,
 about $\sim 10^{-11}$, which can not explain the the discrepancy
 between the experiments and the theoretical prediction.
 Due to the heavy neutrino mass suppression, $\nu_R H^\pm$ loop is even smaller, about $10^{-31}$,
 which is too small and does not show in Fig.\ref{fig2}.

When $m_{\nu_R}$ is very small, the $\nu_R W$ loop contribution given as Eq.(\ref{anu-w1}) is large,
just as that in Ref.\cite{g-2-review2,2109.06089}, $\Delta a_\mu^{\rm \nu_R W}$
approximatively equals to the one-loop contribution predicted by the SM,
 $\Delta a_\mu^{\rm \nu_R W}\sim 10^{-9}$. Note that the contribution is positive.
However in our case, $m_{\nu_R}\sim Y_\mu^2 v^2/m_\nu  \gg m_W$, from Eq. (\ref{anu-w1}),
the $\nu_R W$ contribution can be approximately given as $ \frac{-V^2_\mu m^2_\mu}{16 \pi^2 m^2_W} \sim - 10^{-9}$,
which is negative and widens the discrepancy between theory and experiment.
This further constrains the mixing matrix element $V_\mu$, which will have to be significantly suppressed.
So in our parameter space, the total contribution of the one-loop contribution is negative, which is not
possible to arrive at the required order of the experiments.
\def\figsubcap#1{\par\noindent\centering\footnotesize(#1)}

Hence we can conclude that in the twin Higgs models, the one-loop contribution can not
remedy the discrepancy between the experiments and the
theoretical calculation. 
Since in the two-loop Barr-Zee contribution, the large mass ratio
of quarks to leptons may surpass the phase suppression compared to that at the one-loop level,
and perhaps more parameters contribute to the muon anomalous magnetic moment,
it will be of importance to consider the the two-loop level calculation.



%
%

\section{The analytic expressions for the two-loop muon $g-2$ anomaly and the 125 GeV Higgs global fit in  TH models }
\subsection{ The Analytic Expressions for Two-Loop Barr-Zee Muon $g-2$  }

\begin{figure}[H]
\begin{center}
  \epsfig{file=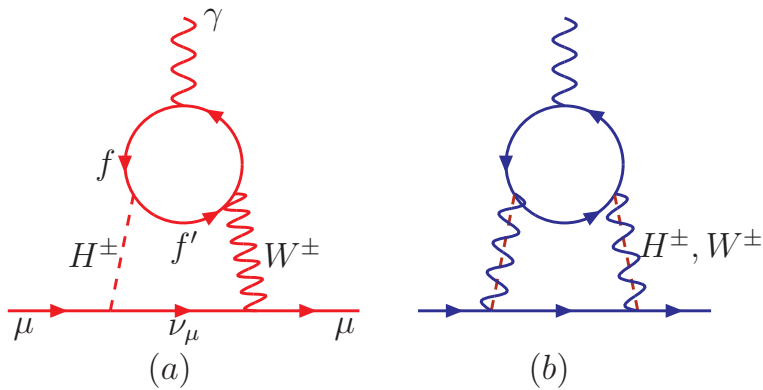,height=5.3cm}
 \vspace{-0.5cm} \caption{ The potential two-loop Barr–Zee muon $g-2$ contributions from the charged Higgs $H^\pm$ and the gauge boson with fermions loop ($fff'$) in TH models, with $f = t, ~b, ~\ell$  and $f' = b, ~t, ~\nu_R$, respectively, where (a) is for one charged Higgs and one gauge boson, and (b) is for either two charged Higgses or two gauge bosons, connecting with the triangle loops. }
\label{fig-2-loop}
\end{center}
\end{figure}

 In TH models, the two-loop Barr-Zee muon $g-2$ anomaly contributions are mediated
 by the charged Higgs $H^\pm$ and the gauge boson with
 the fermions.
The large enhancement factor $m^2_q/m^2_\mu$ may surpass the loop suppression phase space factor $\alpha/\pi$,
so the two-loop contributions could be more important than one-loop ones.
Since the couplings $H^0 \mu \bar \mu$, proportional to $m_\mu/v$,
is too small to have large contribution in the two-loop diagrams,
and due to the discrete symmetry, it is usually difficult to form the couplings such as
 $WZH^+$, $W\gamma H^+$, $Wh^0 H^+$ etc, \cite{1812.08173}, and the heavy gauge boson mass suppression,
 we will just consider the diagrams given in Fig.\ref{fig-2-loop}.

Note that in the Barr-Zee two-loop diagrams there are no two scalars or two $W^\pm$ charged bosons connect to the
triangle loop simultaneously due to the helicity constraints.
Since between the two charged particles in the quark loop, the fermion is the bottom
quark, the slash momentum terms must vanish undergoing a single $\gamma$ matrix because of the
much smaller mass of the bottom quark compared to the top quark, shown as Fig.\ref{fig-2-loop}(b).
Thus, the only contribution comes from Fig.\ref{fig-2-loop}(a).

In Fig.\ref{fig-2-loop} (a),
the fermion loop can consist of not only top and bottom quark $ttb$, $bbt$, but also the lepton with the
right-handed neutrino $\ell\ell v_R$, since the neutrino mass might be quite large.
The Barr-Zee two-loop contribution from quark loop can be written as\cite{g-2-cal,1502.04199,1507.07567,1811.04777,2008.11909},
\bea \nm
   \Delta a_\mu({ttb+bbt})
      &=&
    4m^2_\mu\cdot \frac{g V_{tb}}{\sqrt{2}} \cdot \frac{g }{\sqrt{2}} \cdot \frac{1}{512\pi^4}  \frac{N_t^c }{m_{H^+}^2-m_{W}^2} \\ && \nn
 \int_0^1 dx[Q_tx+Q_b(1-x)]
    \left[ G\left( \frac{m_t^2}{m_{H^+}^2}, \frac{m_b^2}{m_{H^+}^2} \right) - G\left( \frac{m_t^2}{m_W^2}, \frac{m_b^2}{m_W^2} \right) \right]  \\
      &\times & \left( {\Gamma_{tb}^{H^+,R}}^* {\Gamma_{\nu_\mu\mu}^{H^+}} \right) \left[  \frac{m_t}{m_{\mu}} x(1+x) \right]
\label{barr-zee}
\eea
where $\Gamma_{tb}^{H^+,R}=iy_t,~\Gamma_{\nu\mu}^{H^+}=iy_\mu$ are the couplings of $H^+\bar t b$ and $H^+\bar \nu \mu$, respectively,
and $P_{R,L}=\frac{1}{2}(1\pm \gamma^5)$.
The loop function is defined as,
\begin{align}
	G(r^a, r^b) = \frac{\ln\left( \frac{r^a x + r^b(1-x)}{x(1-x)} \right)}{x(1-x) - r^a x - r^b(1-x)}.
\end{align}

For the $\ell\ell \nu_R$ loop when the heavy neutrinos enter into loop, the contribution can be obtained by replacing
$m_t$, $m_b$, $Q_t$ and $Q_b$ by $m_{\nu_R}$, $m_{\ell}$, $Q_{\nu_R}=0$ and $Q_\ell=-1$, respectively,
 in Eq.(\ref{barr-zee}), which can be written explicitly as
\bea \nn
 \Delta a_\mu({\ell\ell\nu_R})&=&4m^2_\mu\cdot \frac{g }{\sqrt{2}} \cdot V_\mu \cdot \frac{1}{512\pi^4}
  \frac{1}{m_{H^+}^2-m_{W}^2} \\ && \nn
 \int_0^1 dx  Q_l(1-x)
    \left[ G\left( \frac{m_{\nu_R}^2}{m_{H^+}^2}, \frac{m_\ell^2}{m_{H^+}^2} \right) - G\left( \frac{m_{\nu_R}^2}{m_W^2}, \frac{m_\ell^2}{m_W^2} \right) \right]  \\
      &\times & \left( {\Gamma_{\nu_R\ell}^{H^+,R}}^* {\Gamma_{\nu_\mu\mu}^{H^+}} \right) \left[  \frac{m_{\nu_R}}{m_{\mu}} x(1+x) \right]
   \label{llvr-loop}
\eea
where $\Gamma_{\nu_R\ell}^{H^+,R}=iY_\mu$.

\subsection{Global Fit of the 125 GeV Higgs}


We will perform a global fit to the 125 GeV Higgs signal data and a large number of observables.
For the given neutral SM-like scalar-field $h$ and its couplings, the $\chi^2_h$ function can be defined as
\begin{align}
\chi^2_h\; =\; \sum_k\; \frac{\left(\mu_k  - \hat{\mu}_k\right)^2}{\sigma_k^2}\, ,
\end{align}
where $k$ runs over the different production(decay) channels considered, and $\mu_k$ is the corresponding theoretical predictions for the TH parameters. $\hat{\mu}_k$ and $\sigma_k$ denote the measured Higgs signal strengths
and their one-sigma errors, respectively, and their choices in this work appear in \cite{sig-streng}, though the data and the references listed are not complete.


The Higgs signal strengths, employed in the experimental data on Higgs searches,
measure the observable cross sections compared to the corresponding SM predictions.
At the LHC, the SM-like Higgs particle is generated by the following
production procedures: gluon fusion ($ g g \rightarrow H$),
vector boson fusion ($ q q^{\prime} \rightarrow q q^{\prime} VV\rightarrow q q^{\prime} H$),
associated production with a vector boson ($q \bar q^{\prime} \rightarrow W H/Z H $),
and the associated production with a $t \bar t$ pair  ($q \bar q/ gg \rightarrow t \bar t  H$).
Meanwhile, the Higgs decay channels are $\gamma \gamma$, $Z Z^{(*)}$, $W W^{(*)}$, $b \bar b$ and $\tau^+ \tau^-$.
The expressions of the Higgs signal strengths have been shown in \cite{1811.04777} and
we will not repeat here.


\section{The Calculation of the Barr-zee Diagrams and the Final Total constraints from the 1- and 2- loop contribution }

Though in Eq.(\ref{lfv-2}) and Eq.(\ref{lfv-1}), the LFV couplings are induced by the SM gauge bosons,
we here still consider the twin, i.e, heavy gauge bosons, which have the same LFV
couplings, but the heavy gauge bosons are quite heavier, about $1$ TeV or more,
just as the discussion above.

\begin{figure}[H]
\begin{center}
\epsfig{file=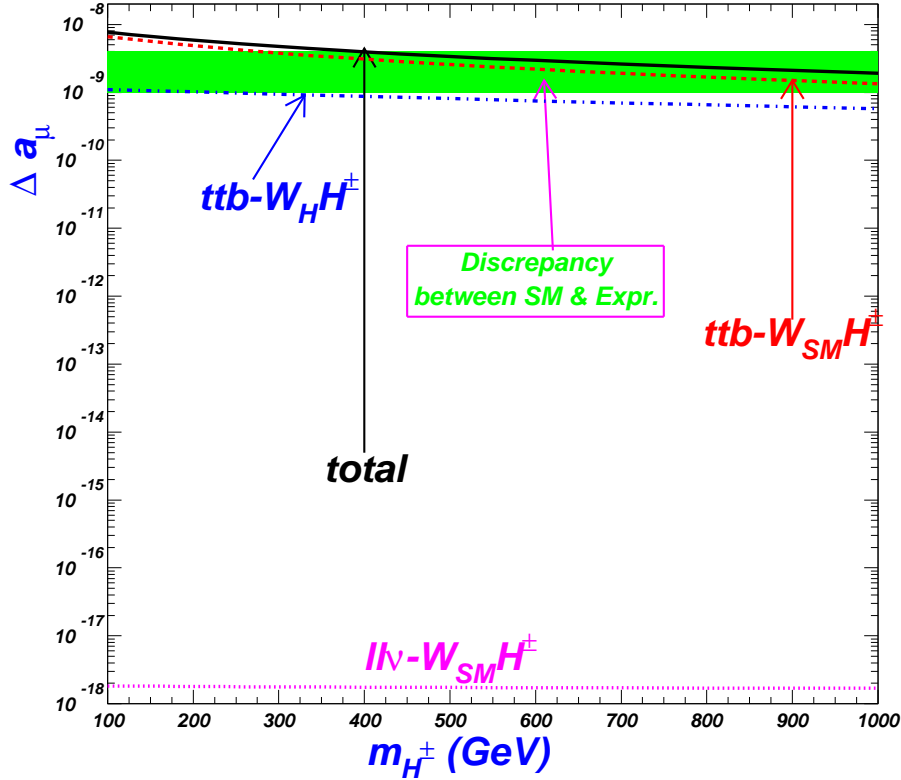,height=10.7cm}
\vspace{-0.5cm} \caption{The comparison among the two-loop $\Delta a_\mu$ contributions of the inner lines of
the charged gauge bosons and the TH heavy charged Higgs boson.
The contribution from the right-handed neutrino loop with the SM charged leptons are also considered.
The green shadow area is the discrepancy between the SM and the measurement for the anomalous magnetic moment $\Delta a_\mu$. }
\label{fig4-comp}
\end{center}
\end{figure}

In Fig.\ref{fig4-comp}, we give the comparison of the contribution at the two-loop level between the inner lines of
the SM charged gauge bosons and the TH heavy charged boson and
that from the right-handed neutrino loop with the SM charged leptons.
The green shadow area in the figure shows the discrepancy between the SM and the measurement for the anomalous magnetic moment $\Delta a_\mu$.

From Fig.\ref{fig4-comp} we can see the heavy mass of the gauge boson suppressing the contribution, and the
contribution is smaller than that of the SM gauge boson, which can be seen from the blue and red curves.
We also consider the lepton $\ell\ell \nu_R$ loop contribution with the charged Higgs and the SM $W^\pm$,
which is shown via the bottom curve in  Fig.\ref{fig4-comp}, and it is much smaller than than the
other contributions.
Hence, the total two-loop contribution comes from not only  the SM  $W^\pm$,
but also the heavy $W^\pm_H$, together with both quark loop and lepton loop,
and not the same as those in the one-loop, the contributions are all positive in the
parameter spaces under consideration.

Note that the coupling $Y_\mu$ is related to the ordinary neutrino masses and  the heavy neutrino mass $m_{\nu_R}$,
just as shown in Sec. III(4), $m_\nu \sim \frac{Y_\mu^2 v^2}{m_{\nu_R}}$, or $m_{\nu_R}\sim \frac{Y_\mu^2 v^2}{m_\nu }$.
Since the upper bound of the heavy neutrino mass is not provided in the experiments, we here just assume the mass
is a parameter determined by the coupling $Y_\mu$ and the ordinary neutrino mass.
With a larger couplings and the tiny neutrino mass, the heavy neutrino mass is quite large,
so the contribution to the two-loop muon anomalous magnetic moment will be very small due to the mass depression, just shown as Fig.\ref{fig4-comp}.

Moreover, from Eq. (\ref{llvr-loop}) and Fig.\ref{fig4-comp}, we know that the contribution from
the $\ell\ell\nu_R$ loop is so small that the only coupling $Y_\mu$ related to it is not
important in determining the $g-2$ calculation,
and at the same time, $Y_\mu$ has nothing to do with other processes.
Hence we here take it as a fixed value: $Y_\mu=0.004$.

\begin{figure}[H]
\begin{center}
\epsfig{file=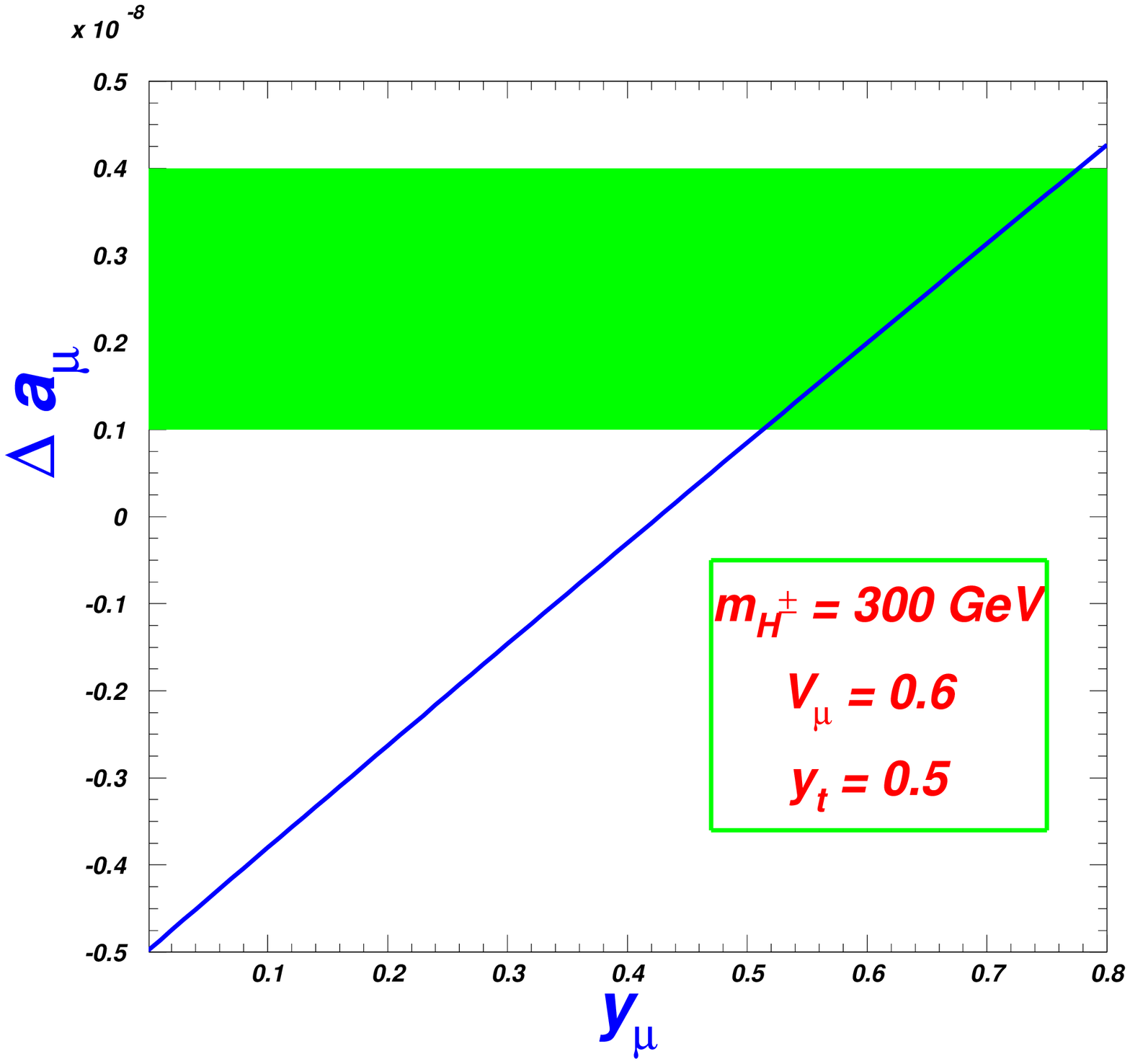,height=8.cm} \hspace{-0.3cm}
\end{center}
\epsfig{file=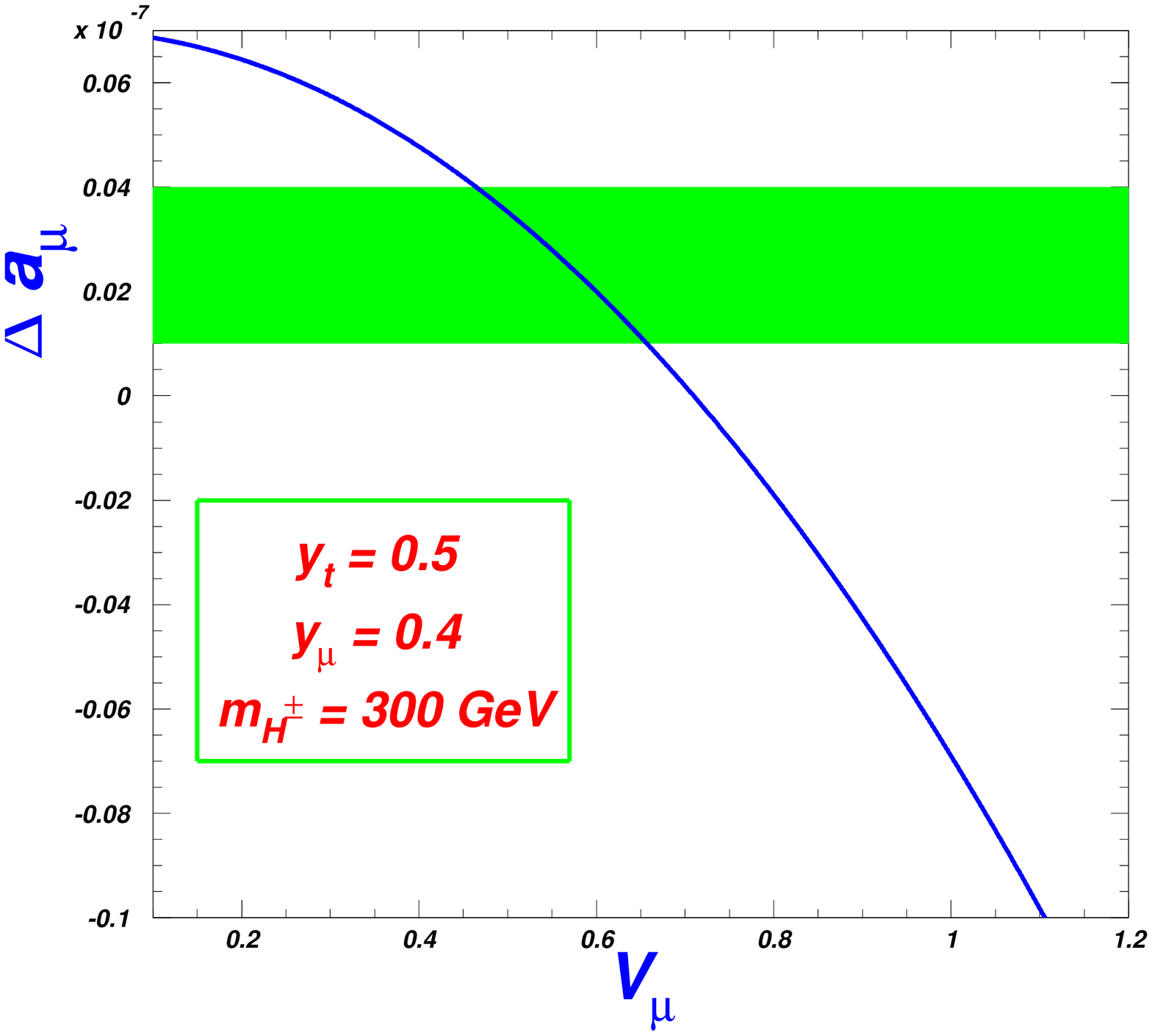,height=8.cm}\hspace{-0.cm}
\epsfig{file=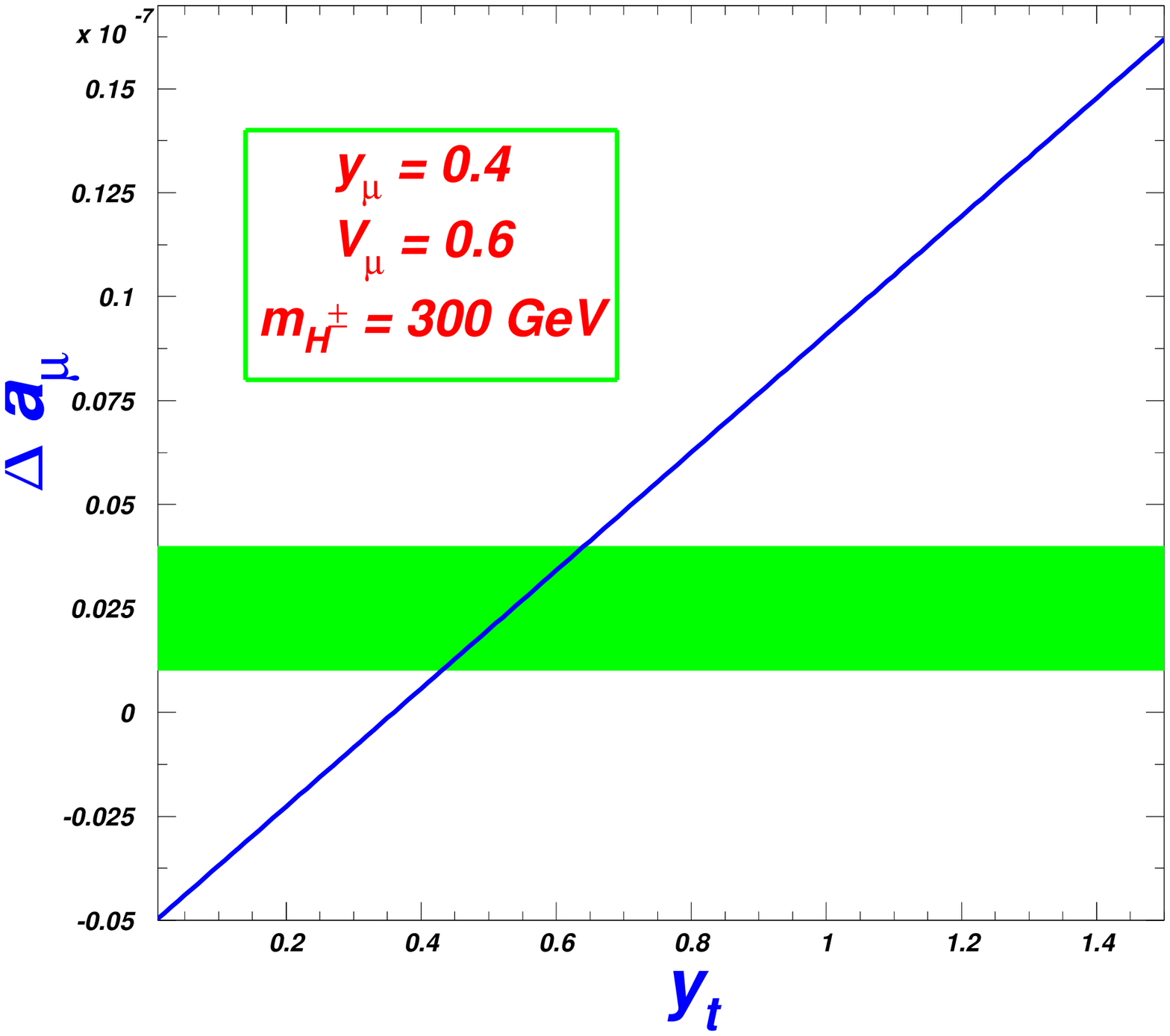,height=8.cm}
\vspace{-1.3cm} \caption{$\Delta a_\mu$ varies with $y_\mu$,
$V_\mu$ and $y_t$ for $m_{H^\pm}=300$ GeV.}
\label{fig-vr-yuka}
\end{figure}
However, the contributions in Fig.\ref{fig4-comp} are given for the parameters only varying with the charged Higgs mass
and without including one-loop ones,
so in Fig.\ref{fig-vr-yuka} we will show the tendency of the total contribution, including not only at the two-loop but also at the one-loop
level, along with the other parameters.
From Fig.\ref{fig-vr-yuka} we can see that $\Delta a_\mu$ increases with the increasing couplings $y_\mu$
 and $y_t$ but decreases with the increasing $V_\mu$, which is because the largest one-loop process to $\Delta a_\mu$
  is from $\nu_R W$ loop with a negative value, and when $V_\mu$ increases, it will be larger towards the negative direction.
Under current parameter conditions, when $V_\mu=0.7$, the total contribution begins to be positive,
and the experiments require $V_\mu$ in the range of $0.47\lesssim V_\mu\lesssim 0.66$.
For $y_\mu$ and $y_t$, to arrive at the discrepancy level, relative large values are required:
$y_\mu \gtrsim 0.52$ and $0.43 \lesssim y_t\lesssim 0.65$.
From Fig.\ref{fig4-comp} and Fig.\ref{fig-vr-yuka} we can see that the total contribution can arrive at the required range in some parameter spaces.


Next, we scan the anomalous magnetic moment and the Higgs global fit
by taking ten thousand random points, so the more the points are left, the higher the
possibility of the event is to be probed in the experiments.
We restate the parameter ranges: $100\leq m_{H^\pm}\leq 1000$ GeV,
$0\leq y_t \leq 1.5 $, $0\leq V_\mu\leq 1.2$ and  $0\leq y_\mu \leq 0.5$.

\begin{figure}[H]
\begin{center}
\epsfig{file=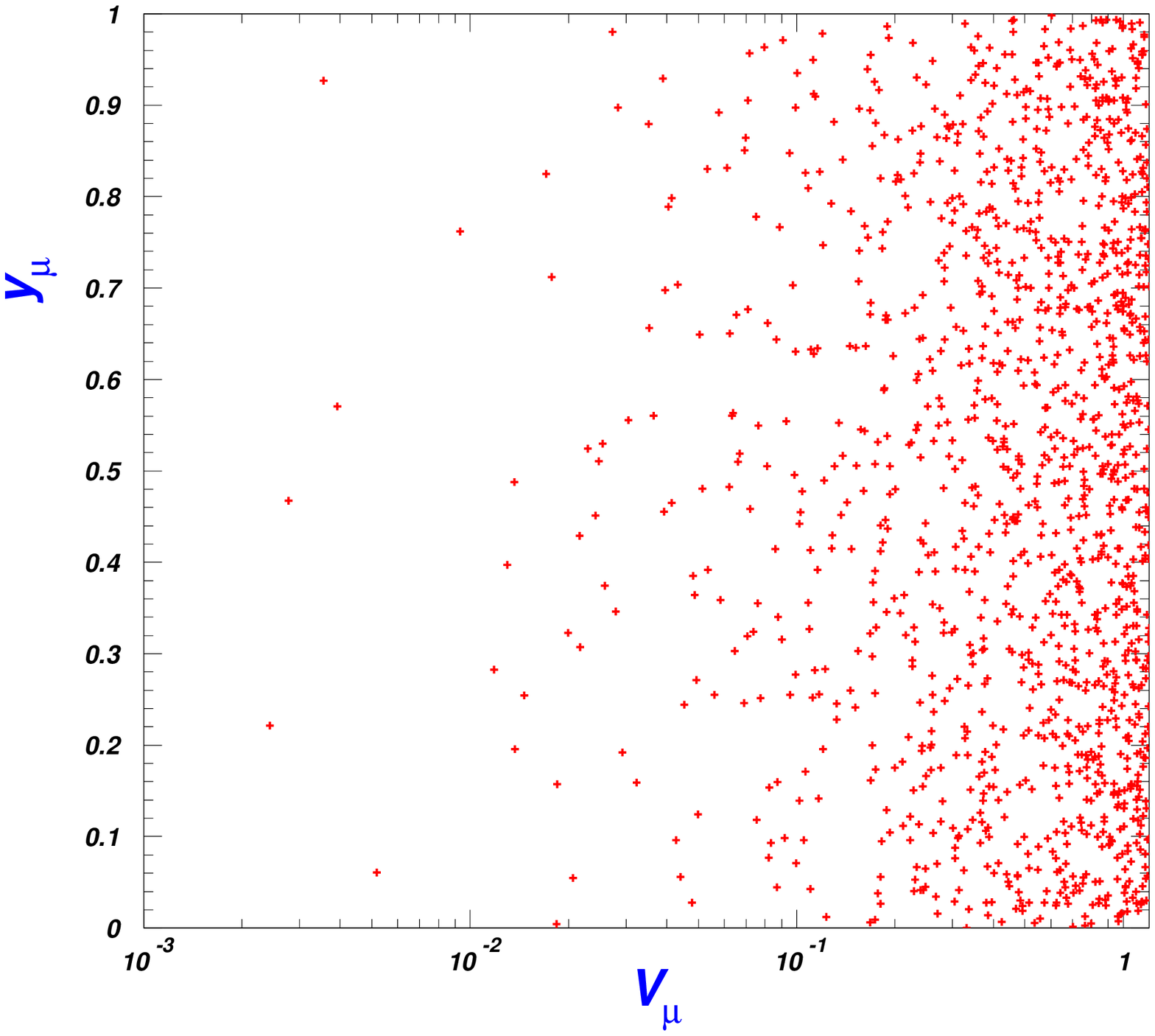,height=8.1cm}\hspace{-0.3cm}
\end{center}
\epsfig{file=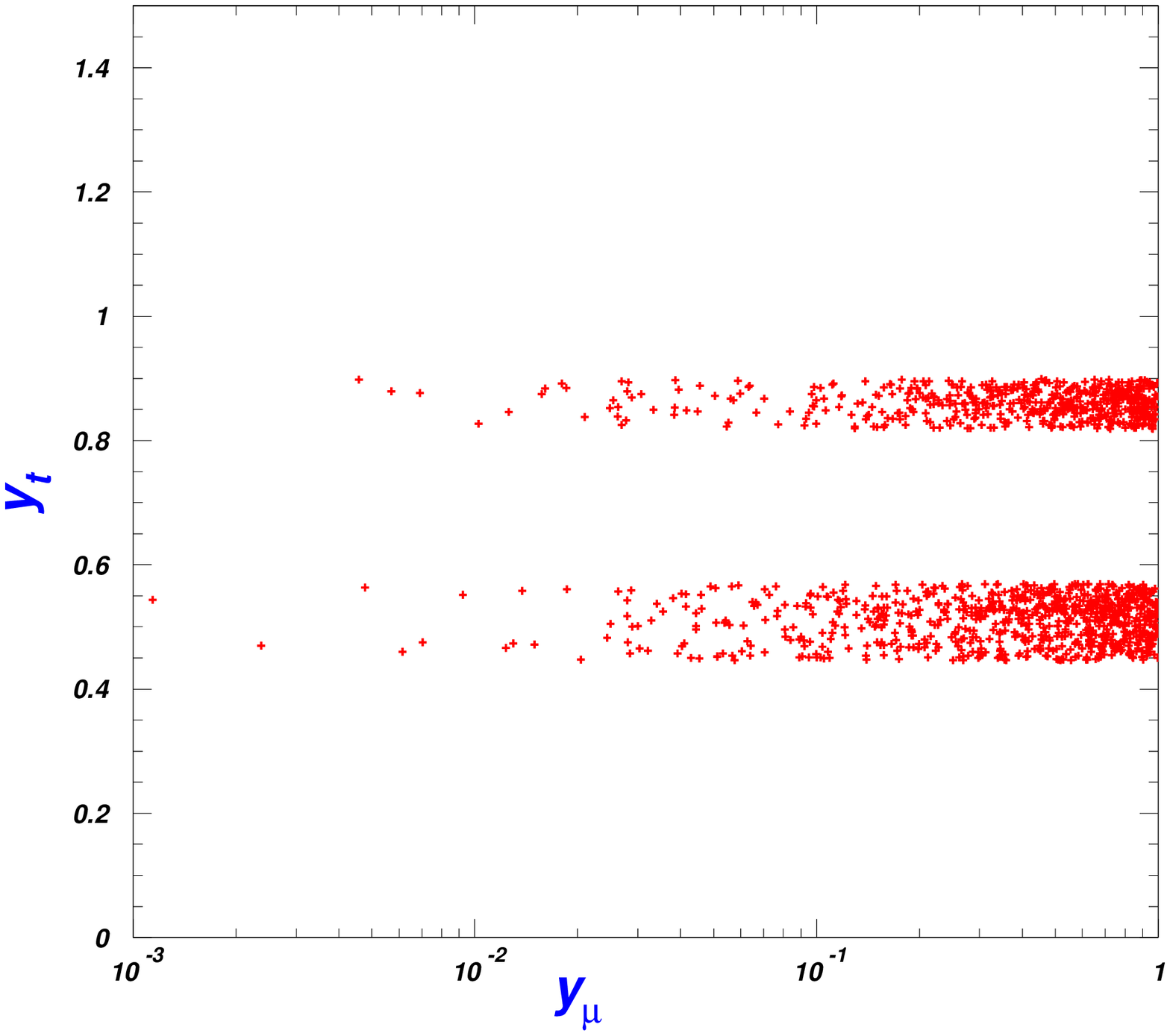,height=8.1cm}\hspace{-0.3cm}
\epsfig{file=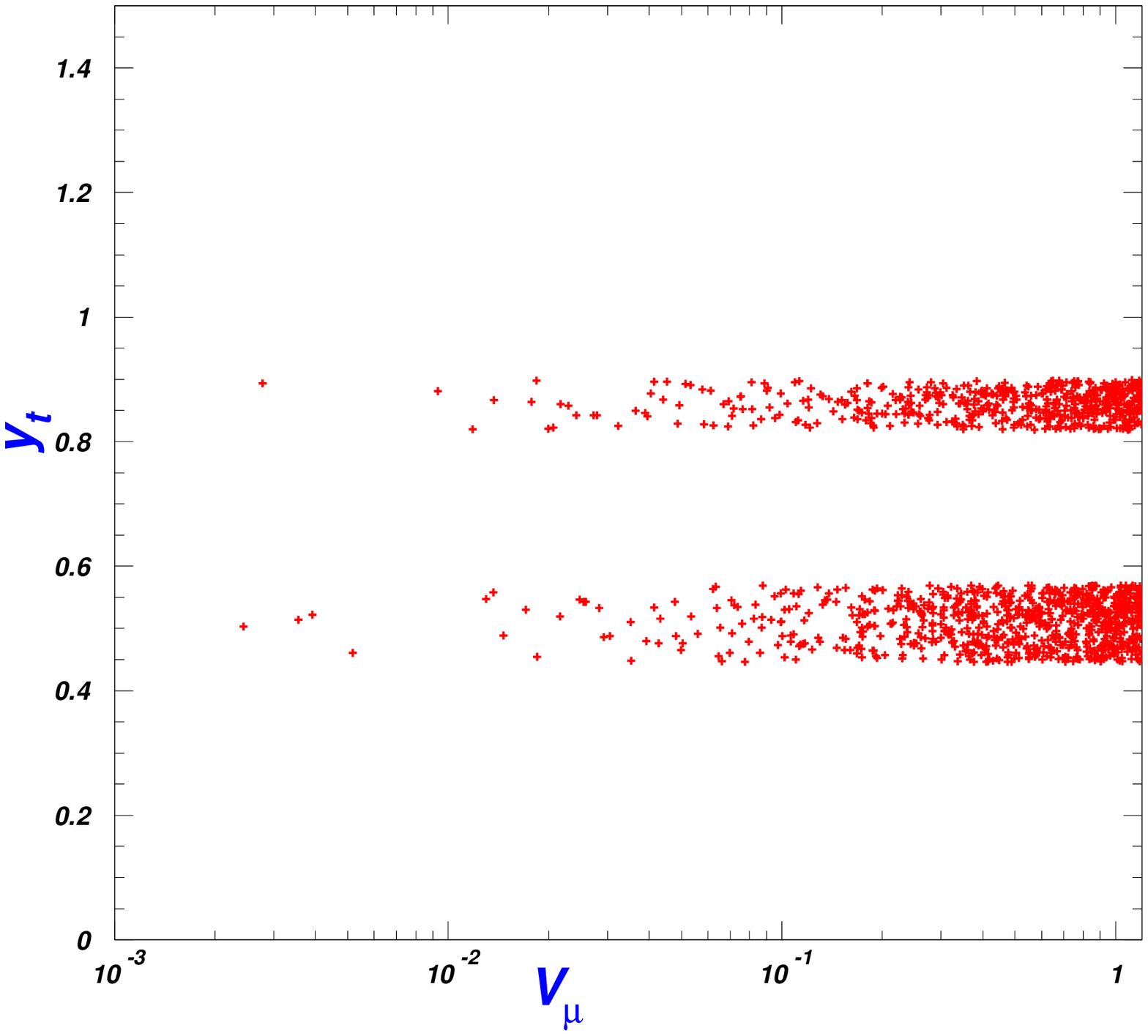,height=8.1cm}
\caption{The surviving samples within 3$\sigma$ ranges of $\chi^2_h$ on the planes
of 
 $V_\mu\sim y_\mu$, $y_\mu \sim y_t$ and  $V_\mu\sim y_t$.}
\label{chi2}
\end{figure}

Since $\chi^2_h$ fits to the 125 GeV Higgs decay signal data,
and we here assume that the neutral Higgs in TH models does not mix with
the charged Higgs, the masses of the charged Higgs and the heavy neutrinos
contribute little to the Higgs strengthen $\chi^2_h$,
and the same time they will hardly be constricted by it, too.
On the other side, due to the quite large top mass,
the top Yukawa coupling contributes to the $\chi^2_h$ value much more than those from others.
So in Fig. \ref{chi2}, we will show the surviving samples only on the planes of the couplings: 
 $y_\mu \sim V_\mu$, $y_t \sim y_\mu $ and  $y_t \sim V_\mu$,
 for the Higgs strengthen in the 3$\sigma$ confidence level.
Fig. \ref{chi2} shows that $\chi^2_h$ favors large $y_t$, while is not sensitive with other couplings.

\begin{figure}[H]
\begin{center}
\hspace{-2cm}
\epsfig{file=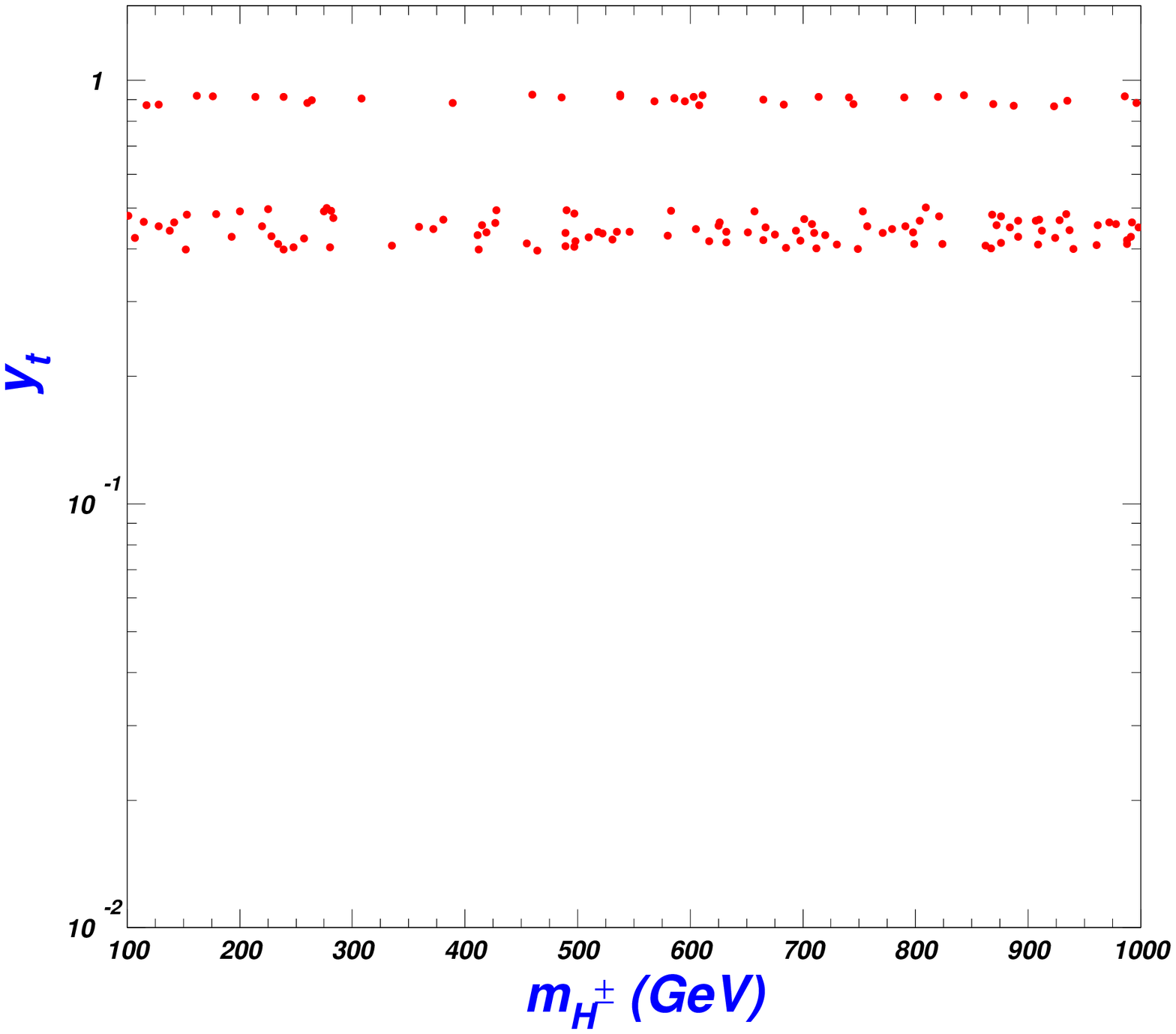,height=7.7cm}
\epsfig{file=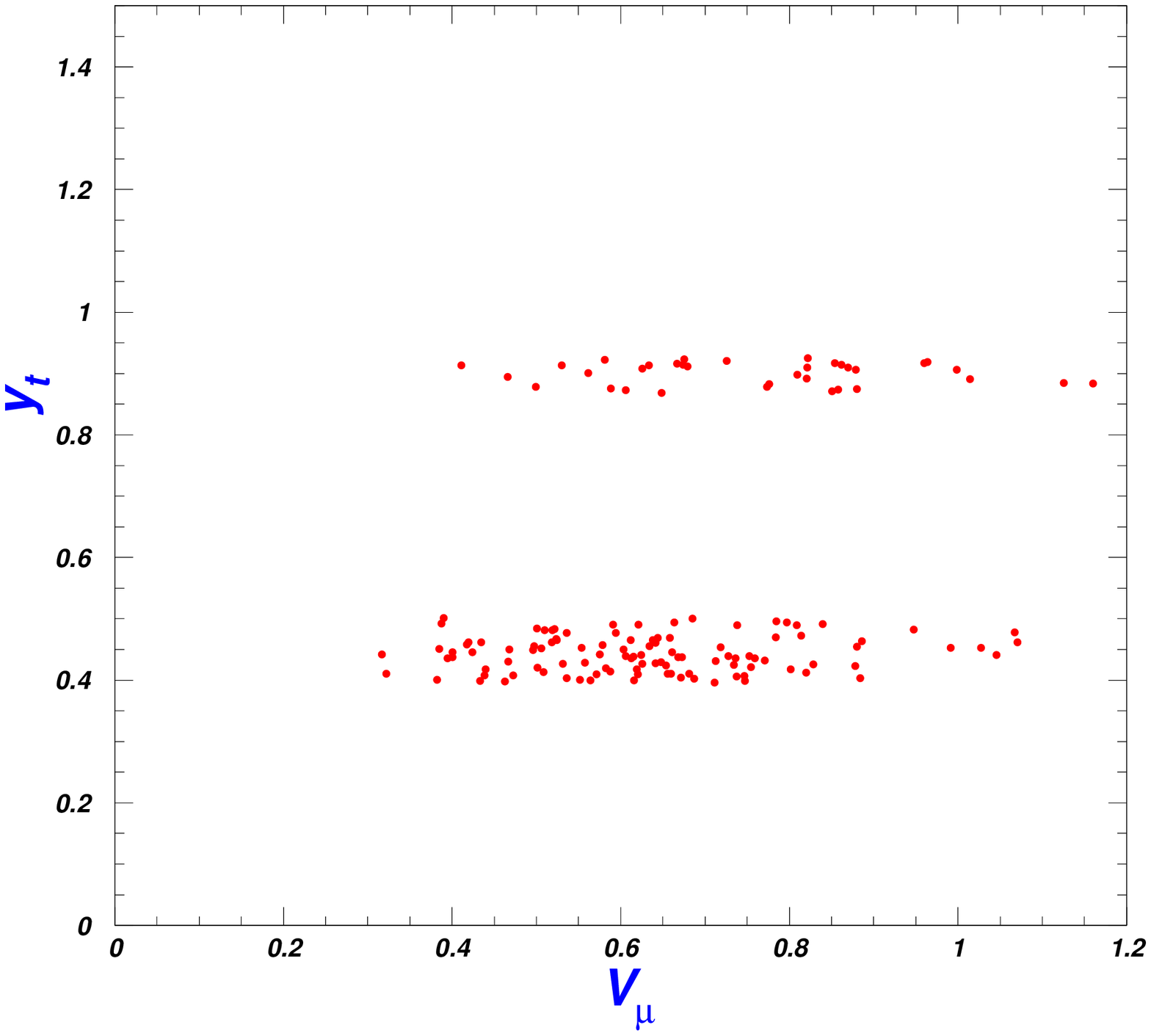,height=7.7cm}\\  \hspace{-2cm}
\epsfig{file=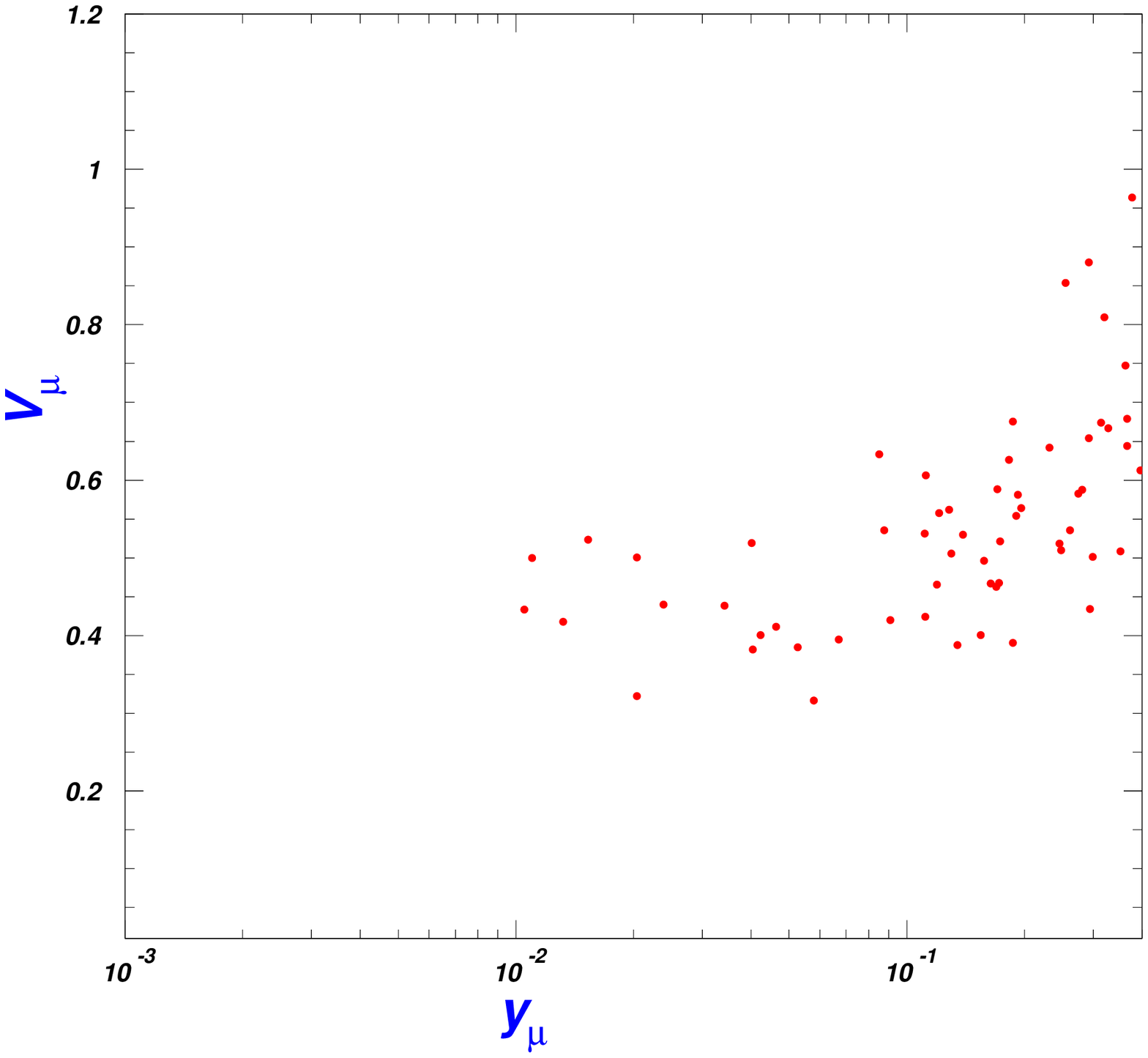,height=7.7cm}
\epsfig{file=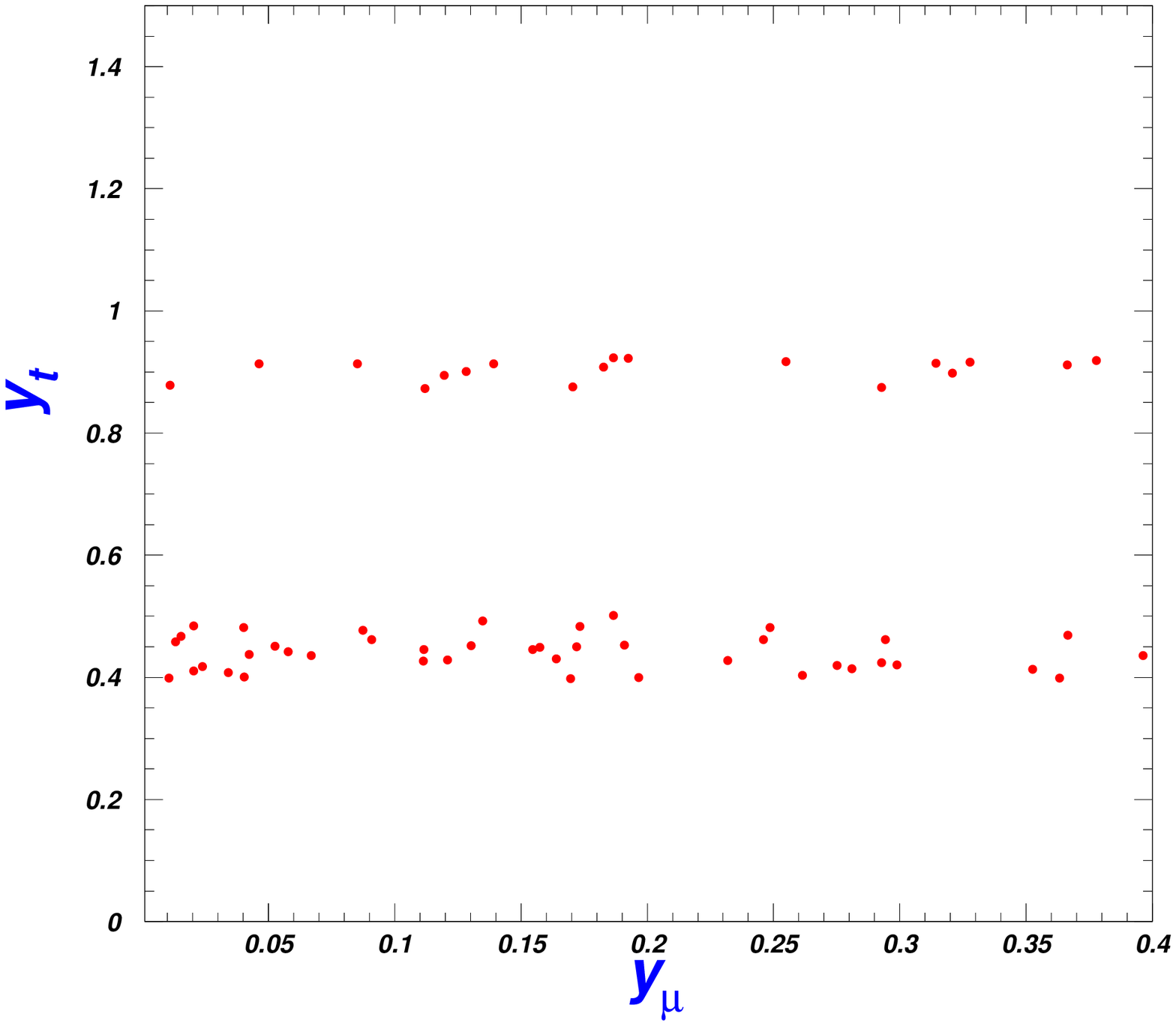,height=7.7cm}
\vspace{-0.25cm} \caption{
The samples satisfying the constraints of Higgs global fit $\chi^2_h$
within 3$\sigma$ range and of the $\Delta a_{\mu}$ from the discrepancy between the experiments and theoretical calculation,
on the planes of $y_t \sim m_{H^\pm}$, $y_t \sim V_\mu$,
$V_\mu \sim y_\mu $ and $y_t\sim y_\mu$. }
\label{chi2-g-2}
\end{center}
\end{figure}

In Fig. \ref{chi2-g-2}, we further impose the two-loop $g-2$ anomaly constraints on the surviving sample points of $\chi^2_h$ within 3$\sigma$
possibility on the planes of $y_t \sim m_{H^\pm}$, $y_t \sim V_\mu$, $V_\mu \sim y_\mu $, and $y_t \sim y_\mu $.
From the first figure in Fig. \ref{chi2-g-2}, we can see that the charged Higgs mass is not constrained too much,
but  the surviving samples favor a larger $y_t$.

We also see that from Eq.(\ref{barr-zee}) that the $g-2$ anomaly varies with the top Yukawa coupling $y_t$,
so coupling $y_t$ should matter much, which can be seen from the other figures in Fig. \ref{chi2-g-2}.
This relevance will constrain $y_t$ greatly, and it can grossly read as $ 0.4\lesssim y_t\lesssim 0.9$.
Another parameter which receive strong constraint is $y_\mu$ in TH models,
and from Eq.(\ref{barr-zee}) we know if the quark loops contribute much,
the $g-2$ anomaly is also relevant with the parameter $y_\mu$ and favors a large value of $y_\mu$.
So we in Fig. \ref{chi2-g-2} give the allowed ranges of the twos with all the constraints,
from which we can grossly get $ 0.12\lesssim y_\mu\lesssim 0.4$.
However, the two-loop $g-2$ anomaly is not sensitive with the coupling $V_\mu$,
and it is mainly contributed by ttb-induced loop, given in Eq.(20), which can be seen in Fig. \ref{fig4-comp}.
Therefore even with the negative one-loop $\nu_R W$-mediated process contribution,
the total $\Delta a_\mu$ does not constrain so stringent as that in the second diagram of Fig.\ref{fig-vr-yuka}: $0.47\lesssim V_\mu\lesssim 0.66$.
For example, in the the right-upper diagram of Fig. \ref{chi2-g-2}, $V_\mu$
can even be larger than $1$ in extreme cases. 



\section{Conclusion}
We consider the charged Higgs contribution to the muon $g-2$ anomaly
in TH models with the joint constraints of the Higgs global fit data.
After imposing various relevant theoretical and experimental constraints,
we perform the scan over the parameter space of this model to identify the ranges in
favor of the muon $g-2$ explanation. We find that the muon $g-2$ anomaly can be explained in TH models
in some parameter spaces. The Higgs direct search limits from LHC contribute most largely in all the constraints.
With the joint constraints from the 125 GeV Higgs signal data, the precision electroweak data,
and the leptonic decay, we find that the muon g-2 anomaly constrains the coupling of charged Higgs to the lepton $ y_\mu$,
 the top Yukawa $y_t$, and the heavy gauge boson coupling to the lepton $V_\mu$
roughly as $ 0.12\lesssim y_\mu\lesssim 0.4$, $0.4\lesssim y_t\lesssim 0.9$,
and $ 0.47\lesssim V_\mu\lesssim 1$.

\section*{Acknowledgment}
This work was supported by the National Natural Science Foundation
of China under grant 12075213,
by the Fundamental Research Cultivation Fund for Young Teachers of Zhengzhou University(JC202041040)
 and the Academic Improvement Project of Zhengzhou University.


\begin{thebibliography}{99}

%


%
\bibitem{fnal-g-2} B. Abi et al. [Muon g-2 Collab.], Phys. Rev. Lett. 126, 141801 (2021), arXiv:2104.03281.


\bibitem{BNL-bennett}P. Zyla et al. [Particle Data Group Collab.], PTEP 2020, 083C01 (2020);
G. W. Bennett et al. [Muon g-2 Collab.], \PRD 73, (2006) 072003, arXiv:hep-ex/0602035.


\bibitem{g-2-SM} P. Athron, C. Bal$\acute{a}$zs, D. H. Jacob, W. Kotlarski, D. St$\ddot{o}$ckinger and H. St$\ddot{o}$ckinger-Kim,
JHEP09, (2021) 080, arXiv:2104.03691;
T. Aoyama et al., 
Phys. Rept. 887 (2020) 1-166, arXiv:2006.04822.
%
%
\bibitem{g-2-review1} J. P. Miller, E. de Rafael, and B. L. Roberts, Rept.
Prog. Phys. 70, (2007) 795, arXiv: hep-ph/0703049.
\bibitem{g-2-review2} F. Jegerlehner and A. Nyffeler, Phys. Rept. 477, (2009) 1
, arXiv:0902.3360. 

\bibitem{0902.3360}
F. Jegerlehner, A. Nyffeler, Phys. Rept. 477 (2009) 1-110, ArXiv: 0902.3360.

%
\bibitem{g-2-review4-1610.06587} M. Lindner, M. Platscher, F. S. Queiroz, Phys.
Rept. 731, (2018) 1, arXiv:1610.06587.
%
\bibitem{g-2-review5} F. Jegerlehner,
 Acta Phys.Polon.B 49 (2018) 1157, arXiv:1804.07409.  
%
\bibitem{2002.12347}
Sz. Borsanyi, Z. Fodor, J.N. Guenther, C. Hoelbling, S.D. Katz et al., Nature 593 (2021) 7857, 51, arXiv: 2002.12347.

\bibitem{1811.04777}Guo-Li Liu, Qing-Guo Zeng,
\EPJC79, 612 (2019), arXiv:1811.04777.
%
\bibitem{2106.04466}  
Zhuang Li, Guo-Li Liu, Fei Wang, Jin Min Yang, Yang Zhang,
 JHEP12(2021)219, arXiv:2106.04466.
 %

\bibitem{litt-hier1} Z. Chacko, H.-S. Goh, and R. Harnik, Phys. Rev. Lett. 96, 231802 (2006), arXiv: hep-ph/0506256.
%

\bibitem{arXiv:hep-ph/0007265}Riccardo Barbieri, Alessandro Strumia, arXiv:hep-ph/0007265.  
%
\bibitem{1501.07890} M. Low, A. Tesi, L. Wang, \PRD91, 095012 (2015), arXiv:1501.07890.

\bibitem{1905.02203} 
J. Serra, S. Stelzl, R. Torre, A. Weiler, JHEP10(2019)060, arXiv:1905.02203.

\bibitem{neff_exp1} R. H. Cyburt, B. D. Fields, K. A. Olive, and T.-H. Yeh, Rev. Mod. Phys. 88, (2016) 015004, arXiv:1505.01076.
\bibitem{neff_exp2} N. Aghanim et al. (Planck Collaboration), A$\&$A 641, (2020)A6, arXiv:1807.06209.
%
\bibitem{1905.08798}
Keisuke Harigaya, Robert McGehee, Hitoshi Murayama, Katelin Schutz, JHEP05, (2020) 155,
arXiv:1905.08798.
\bibitem{1611.07975-neff_mth} Z. Chacko, N. Craig, P. J. Fox, and R. Harnik, JHEP 07, (2017) 023, arXiv:1611.07975.

\bibitem{modif-mth1} N. Craig, A. Katz, M. Strassler, and R. Sundrum, JHEP 07, (2015) 105, arXiv:1501.05310.
 \bibitem{modif-mth2}R. Barbieri, L. J. Hall, and K. Harigaya, JHEP 11, (2016) 172, arXiv:1609.05589.
\bibitem{modif-mth3-1703.06884}C. Csaki, E. Kuflik, and S. Lombardo, Phys. Rev. D 96, (2017) 055013, arXiv:1703.06884.
%
\bibitem{modif-mth4} B. Batell and C. B. Verhaaren, JHEP 1912, (2019) 010, arXiv:1904.10468.
%
 \bibitem{modif-mth5-1905.00861}D. Liu and N. Weiner, arXiv:1905.00861.




\bibitem{modif-mth6} N. Craig, S. Koren, and T. Trott, JHEP 05, (2017) 038, arXiv:1611.07977.
 \bibitem{modif-mth7}N. Craig, S. Knapen, P. Longhi, and M. Strassler, JHEP 07, (2016) 002, arXiv:1601.07181.

\bibitem{type I}
 P. Minkowski, Phys. Lett. B67, (1977) 421;\\
 R. N. Mohapatra and G. Senjanovic, Phys. Rev. Lett. 44, (1980) 912;\\
 T. Yanagida, Conf. Proc. C 7902131 (1979)95;\\
 M. Gell-Mann, P. Ramond, and R. Slansky, Conf. Proc. C 790927, (1979) 315.

\bibitem{barr-zee}S. M. Barr and A. Zee, Phys. Rev. Lett. 65 (1990) 21 [Erratum-ibid. 65 (1990) 2920]

\bibitem{0909.5148}
Junjie Cao, Peihua Wan, Lei Wu, Jin Min Yang, \PRD80, 071701, (2009), arXiv:0909.5148.
\bibitem{1912.10225}
Junjie Cao, Jingwei Lian, Lei Meng, Yuanfang Yue, Pengxuan Zhu, Phys. Rev. D 101, 095009 (2020), arXiv:1912.10225.
%

\bibitem{hier_prob1} P. Fayet, Phys. Lett. B,69 (1977)489.
\bibitem{hier_prob2} S. Dimopoulos and H. Georgi, Nucl. Phys. B193, (1981 150).
\bibitem{hier_prob3} N. Arkani-Hamed, A. G. Cohen, and H. Georgi, Phys. Lett. B513, (2001) 232, arXiv: hep-ph/0105239.
%
\bibitem{susy-pheno}H. Murayama, 
Rpt. no.: UCB-PTH-00/05, arXiv: hep-ph/0002232.
\bibitem{susy1} L. Maiani, 
Conf. Proc.C 7909031, (1979) 1.
\bibitem{compo-scale} D. B. Kaplan and H. Georgi, Phys. Lett. B136, (1984) 183-186; 
D. B. Kaplan, H. Georgi, and S. Dimopoulos, Phys. Lett. B136, (1984) 187-190. 


\bibitem{tc-review}
  S. Weinberg, Phys. Rev. D13, (1976) 974; ibid, D19, (1979) 1277;
  L. Susskind, Phys. Rev. D20, (1979) 2619;
  E. Farhi, L. Susskind, Phys. Rept. 74, (1981) 277,
C. T. Hill, \PLB345, 483 (1995), arXiv: hep-ph9411426;
            K. Lane and E. Eichten, \PLB352, 383 (1995), arXiv: hep-ph/9503433;
            K. Lane, \PLB483, 96 (1998), arXiv: hep-ph/9805254;
            G. Cvetic,\RMP71, 513 (1999), arXiv: hep-ph/9702381;
            C.~T.~Hill and E.~H.~Simmons,  
            Phys. Rept.{\bf 381}, (2003) 235-402; Erratum-ibid.390, (2004) 553, arXiv: hep-ph/0203079.
\bibitem{parti-tev}ATLAS Collaboration, M. Aaboud et al., 
Phys. Rev. D98 (2018), 032008, arXiv:1803.10178;
 CMS Collaboration, A. M. Sirunyan et al.,
JHEP 05 (2018) 025, arXiv:1802.02110;
G. Aad et al. [ATLAS Collaboration],
Phys. Lett. B758 (2016) 249-268, arXiv:1602.06034;
 CMS Collaboration, A. M. Sirunyan et al.,
JHEP 08 (2018) 177, arXiv:1805.04758.
%

%
\bibitem{litt-hier2} R. Barbieri, T. Gregoire, and L. J. Hall,
Rpt. no.: CERN-PH-TH/2005-162, UCB-PTH-05/25, LBNL-58803, arXiv: hep-ph/0509242. 
\bibitem{litt-hier3} Z. Chacko, Y. Nomura, M. Papucci, and G. Perez, JHEP 01, 126 (2006), arXiv: hep-ph/0510273. 
\bibitem{litt-hier4} G. Burdman, Z. Chacko, H.-S. Goh, and R. Harnik, JHEP 02, 009 (2007), arXiv: hep-ph/0609152. 
\bibitem{litt-hier5} H. Cai, H.-C. Cheng, and J. Terning, JHEP 05, 045 (2009), arXiv:0812.0843. 
\bibitem{litt-hier6} D. Poland and J. Thaler, JHEP 11, 083 (2008), arXiv:0808.1290. 
\bibitem{litt-hier7} B. Batell and M. McCullough, Phys. Rev. D92, 073018 (2015), arXiv:1504.04016.
\bibitem{litt-hier8} J. Serra and R. Torre, Phys. Rev. D 97, (2018) 035017, arXiv:1709.05399. 
\bibitem{litt-hier9} Csaba Cs$\acute{a}$ki, T. Ma, and J. Shu, Phys. Rev. Lett. 121, (2018) 231801, arXiv:1709.08636.
%
\bibitem{1711.05300}Zackaria Chacko, Can Kilic, Saereh Najjari, Christopher B. Verhaaren,
Phys. Rev. D 97, 055031 (2018), arXiv:1711.05300.

%
\bibitem{Barbieri:2015lqa} R. Barbieri, D. Greco, R. Rattazzi, and A. Wulzer, 
	JHEP08, (2015) 161, arXiv:1501.07803.
\bibitem{Batra:2008jy} P. Batra and Z. Chacko, 
Phys. Rev. D79 (2009) 095012, arXiv:0811.0394.
%
\bibitem{1702.04399}
Meziane Chekkal, Amine Ahriche, Amine Bouziane Hammou, Salah Nasri,
\PRD 95, (2017) 095025, arXiv:1702.04399.
%
\bibitem{2202.01228}Gegenbauer's Twin
Gauthier Durieux, Matthew McCullough, Ennio Salvioni, JHEP 05 (2022) 140, 	arXiv:2202.01228.
%
\bibitem{Emam:2007dy} W.~Emam and S.~Khalil,
  Eur.\ Phys.\ J.\  C {\bf 55}, 625 (2007), arXiv:0704.1395.
\bibitem{1105.1047}
W. Abdallah, A. Awad, S. Khalil, H. Okada, Eur. Phys. J. C 72, (2012) 2108, arXiv:1105.1047.
%
\bibitem{mns-maki-1962}   Z.~Maki, M.~Nakagawa and S.~Sakata,
  Prog.\ Theor.\ Phys.\  {\bf 28}, (1962) 870.
%
\bibitem{0712.4019}
A.G. Akeroyd, Mayumi Aoki, Hiroaki Sugiyama, Phys. Rev. D77, (2008) 0750108, arXiv:0712.4019.
%
\bibitem{charged-neutral-mass} D. Chowdhury and O. Eberhardt, JHEP 05, 161 (2018), arXiv: 1711.02095;
Qing-Hong Cao, Hao-Lin Li, Ling-Xiao Xu, Jiang-Hao Yu,
arXiv: 2107.08343. 

\bibitem{0611015-su} 
Hock-Seng Goh, Shufang Su, \PRD75, (2007) 075010, arXiv: hep-ph/0611015.
%
\bibitem{MEG-2013-2016} MEG Collaboration, J. Adam et al., 
\PRL 110, (2013) 201801, arXiv:1303.0754;
MEG Collaboration, A. M. Baldini et al.,
\EPJC76, (2016) 434, arXiv:1605.05081.
%
\bibitem{gau-bo-mass-1} G. Aad et al. [ATLAS Collaboration], 
Phys. Lett. B 796, (2019) 68, arXiv:1903.06248.
\bibitem{gau-bo-mass-2}
                      W. Adam et al. [CMS Collaboration], JHEP 07 (2021) 208, arXiv:2103.02708.
\bibitem{gau-bo-mass-3} G. Aad et al. [ATLAS Collaboration], 
\PRD 100, 052013 (2019),  arXiv:1906.05609.  
%
\bibitem{pdg-2018} M. Tanabashi et al. [Particle Data Group], 
Phys. Rev. D 98, 030001 (2018). 
\bibitem{ssm-1989} G. Altarelli, B. Mele and M. Ruiz-Altaba,
Z. Phys. C 45, 109 (1989) Erratum: Z. Phys. C 47, 676 (1990).
%
\bibitem{1912.02106}
A. Pankov, P. Osland, I.  Serenkova and V. Bednyakov, \EPJC80, (2020) 503, arXiv:1912.02106.
%
\bibitem{1loop-deltaamu}
J. P. Leveille, Nucl. Phys. B 137, (1978) 63;
S. R. Moore, K. Whisnant, and Bing-Lin Young, \PRD31, (1985) 105;
Farinaldo S. Queiroz, William Shepherd, \PRD89 (2014) 095024, arXiv:1403.2309.
%

%
\bibitem{2109.06089} 
L. T. Hue, K. H. Phan, T. P. Nguyen, H. N. Long, H. T. Hung, e-Print: 2109.06089.
%
\bibitem{g-2-cal}
 A. Broggio, E. J. Chun, M. Passera, K. M. Patel and S. K. Vempati, JHEP 1411 (2014) 058, arXiv:1409.3199;
 L. Wang and X. F. Han, \JHEP05, 039 (2015), arXiv:1412.4874;
 A. Dedes and H. E. Haber, \JHEP0105 (2001) 006, arXiv: hep-ph/0102297;
 J. F. Gunion, \JHEP0908, (2009) 032, arXiv:0808.2509;
 K. M. Cheung, C. H. Chou and O. C. W. Kong, \PRD64, (2001) 111301, arXiv: hep-ph/0103183;
 D. Chang, W. F. Chang, C. H. Chou and W. Y. Keung, \PRD63 (2001) 091301, arXiv: hep-ph/0009292;
 M. Krawczyk, Acta Phys. Polon. B33, (2002) 2621, arXiv: hep-ph/0208076;
 F. Larios, G. Tavares-Velasco and C. P. Yuan, \PRD64, (2001) 055004, arXiv: hep-ph/0103292;
 K. Cheung and O. C. W. Kong, \PRD68, (2003) 053003, arXiv: hep-ph/0302111;
 A. Arhrib and S. Baek, \PRD65, (2002) 075002, arXiv: hep-ph/0104225;
 S. Heinemeyer, D. Stockinger and G. Weiglein, \NPB690, (2004) 62, arXiv: hep-ph/0312264;
 O. C. W. Kong, arXiv: hep-ph/0402010; 
 K. Cheung, O. C. W. Kong and J. S. Lee, \JHEP0906, (2009) 020, arXiv:0904.4352.
 \bibitem{1812.08173}See e.g,
Can Kilic, Saereh Najjari, Christopher B. Verhaaren,
Phys. Rev. D 99, 075029 (2019), arXiv:1812.08173, doi10.1103/PhysRevD.99.075029.
\bibitem{1502.04199}
Victor Ilisie, \JHEP04, (2015) 077, arXiv:1502.04199.
\bibitem{1507.07567}
A. Crivellin, J. Heeck, P. Stoffer, \PRL116,(2016) 081801, arXiv:1507.07567.
%
\bibitem{2008.11909}
Mariana Frank, Ipsita Saha, Phys. Rev. D 102, (2020) 115034, arXiv: 2008.11909.
\bibitem{sig-streng}
 ATLAS Collaboration, \PRD98, (2018) 052003, arXiv:1807.08639;
%
 CMS Collaboration,	\PLB780, (2018) 501, arXiv:1709.07497;
%
CMS Collaboration,	\JHEP03, (2019) 026, arXiv:1804.03682.
%

\bibitem{1006.5534} 
Liang Han, Xiao-Gang He, Wen-Gan Ma, Shao-Ming Wang, Ren-You Zhang, JHEP 09 (2010) 023, arXiv: 1006.5534.

\end{thebibliography}
\end{document}